\makeatletter
\@ifundefined{@parse@version@dash}{%
	\def\@parse@version#1{\@parse@version@0#1}
	\def\@parse@version@#1/#2/#3#4#5\@nil{%
		\@parse@version@dash#1-#2-#3#4\@nil}
	\def\@parse@version@dash#1-#2-#3#4#5\@nil{%
		\if\relax#2\relax\else#1\fi#2#3#4 }
}{}
\makeatother

\documentclass[%
reprint,
superscriptaddress,
amsmath,amssymb,
prl,
showkeys
]{revtex4-2}

\usepackage{color}
\usepackage{graphicx}
\usepackage{dcolumn}
\usepackage{bm}
\usepackage{hyperref}
\newcommand*{\dif}{\mathop{}\!\mathrm{d}}

\newtheorem{theorem}{Theorem}

\newtheorem{corollary}{Corollary}

\begin{document}

\title{Theorem on the Compatibility of Spherical Kirigami Tessellations}

\author{Xiangxin Dang}
\affiliation{State Key Laboratory for Turbulence and Complex Systems, Department of Mechanics and Engineering Science, College of Engineering, Peking University, Beijing 100871, China}
\author{Fan Feng}
\affiliation{Cavendish Laboratory, University of Cambridge, Cambridge CB3 0HE, United Kingdom}
\author{Huiling Duan}
\author{Jianxiang Wang}
\email{jxwang@pku.edu.cn}
\affiliation{State Key Laboratory for Turbulence and Complex Systems, Department of Mechanics and Engineering Science, College of Engineering, Peking University, Beijing 100871, China}
\affiliation{CAPT-HEDPS, and IFSA Collaborative Innovation Center of MoE, College of Engineering, Peking University, Beijing 100871, China}

\date{\today}

\begin{abstract}
We present a theorem on the compatibility upon deployment of kirigami tessellations restricted on a spherical surface with patterned slits forming freeform quadrilateral meshes. We show that the spherical kirigami tessellations have either one or two compatible states, i.e., there are at most two isolated strain-free configurations along the deployment path. 
The theorem further reveals that the rigid-to-floppy transition from spherical to planar kirigami tessellations is possible if and only if the slits form parallelogram voids along with vanishing Gaussian curvature, which is also confirmed by an energy analysis and simulations. On the application side, we show a design of bistable spherical dome-like structure based on the theorem. Our study provides new insights into the rational design of morphable structures based on Euclidean and non-Euclidean geometries.
\end{abstract}


\maketitle


\emph{Introduction.} Prescribed cuts on kirigami structures can induce desired deformations across various scales \cite{blees2015graphene, Zhang2015A, tang2017Design, xu2017origami, callens2018flat, Hanakata2018Accelerated, rafsanjani2019propagation, Guo2020Designing}. This concept has been utilized to design metamaterials \cite{rafsanjani2016bistable, rafsanjani2017Buckling, Yi2018Multistable, Tang2019Programmable, an2020Programmable}, morphable structures \cite{Cho2014Engineering, Sussman2015Algorithmic, celli2018shape, choi2019programming, choi2021compact, jin2020kirigami}, nanocomposites \cite{shyu2015kirigami}, soft robotics \cite{rafsanjani2018kirigami}, and mechanical actuators \cite{dias2017kirigami}.
In these applications, deployability and multistability play significant roles in determining the energy landscapes and morphing routes of the deploying process.
The deformation of deployable (rigidly deployable, to be exact) kirigami structures can be idealized as continuous rotations of rigid panels connected by flexible hinges at the corner \cite{choi2021compact}.
Thus, from the viewpoint of geometry, rigid deployability means having a series of piecewise isometric transformations between the undeployed configuration and any deployed configurations along the path of deployment.
{ By contrast}, multistability emerges if piecewise isometric transformations only { exist} at a finite number of states on the deployment path \cite{rafsanjani2016bistable, choi2019programming}, { which is physically equivalent to discontinuous connections between stress-free configurations} \footnote{{ Here we only consider stress-free stable configurations, which are determined solely by geometry.}}.
Existence of such isometric transformations is also referred to as \emph{compatibility}, a term originally proposed to design rigidly deployable origami tessellations \cite{BELCASTRO2002273, tachi2009generalization, feng2020designs}.

\begin{figure}[!tp]
	\centering
	\includegraphics[width=8.6 cm]{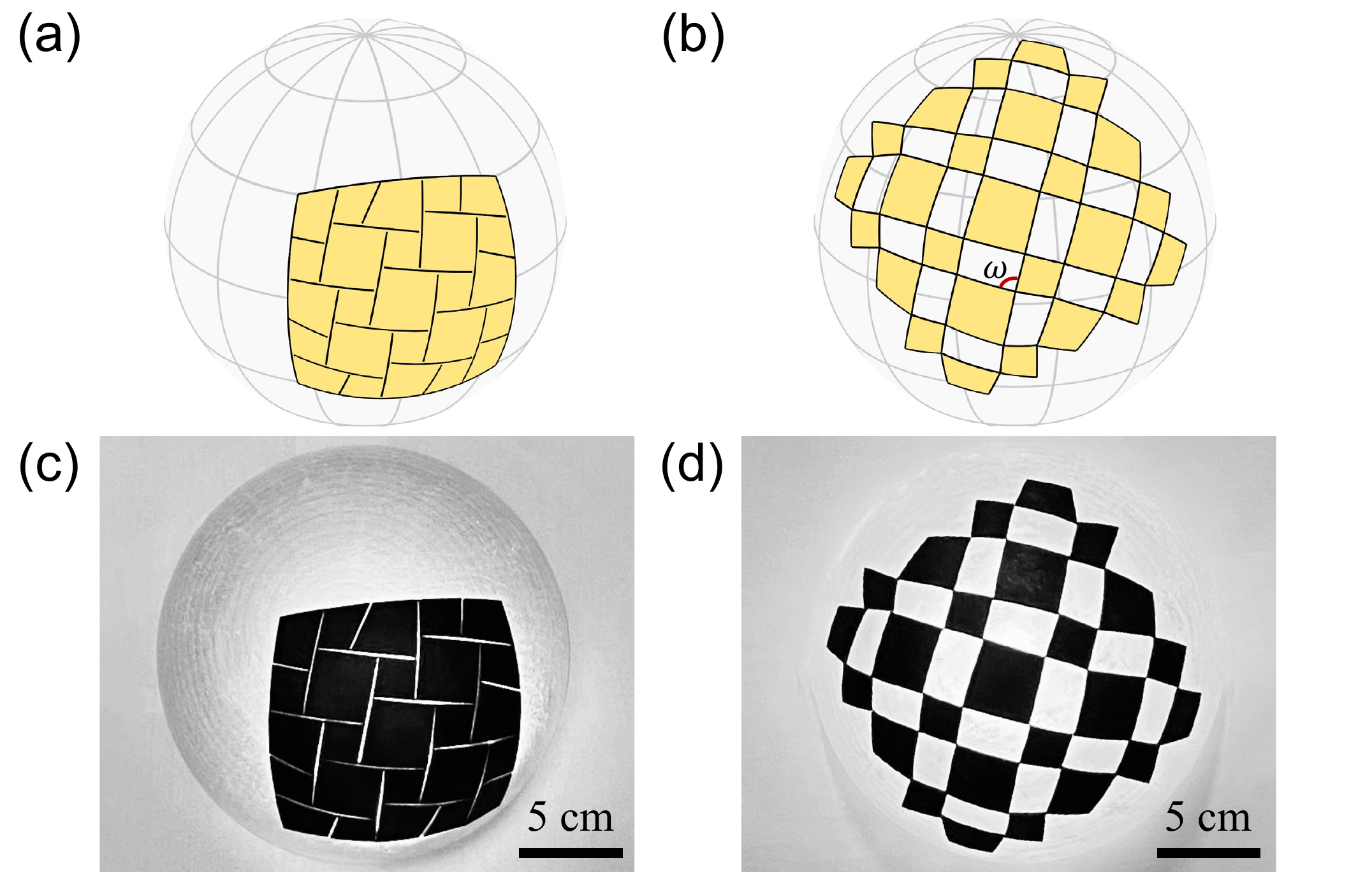}
	\caption{$5\times5$ SQK tessellation. (a) The kirigami pattern. (b) The compatible configuration at $\omega=\pi/2$. (c) and (d) The physical model made of rubber.}
	\label{fig:intro}
\end{figure}

In order to achieve desired energy landscapes with deployability or multistability upon deployment of kirigami structures, pioneering studies on geometrical and topological design principles have been carried out \cite{Castle2014Making, Castle2016Additive, chen2016topological, yang2018geometry, chen2020deterministic, Wang2020Keeping, choi2019programming, choi2021compact, Jiang2020Freeform}.
However, most existing works focus on classical patterns with planar symmetry \cite{grunbaum1987tilings}, e.g., the well-known rotating squares \cite{grima2000auxetic} and kagome patterns \cite{Sun2012Surface}, while freeform slit distributions can greatly expand the configuration space of kirigami structures \cite{choi2019programming, choi2021compact, Jiang2020Freeform}.
Besides, nearly all the current works on morphable kirigami consider cutting flat sheets to engineer the deployed shapes in two or three dimensions, while very few are focused on cutting curved surfaces of non-Euclidean geometry.
As an example of kirigami on curved developable surfaces, cylindrical shells with prescribed slits have been found to have unusual energy barriers with pop-up deformations compared to flat sheets \cite{rafsanjani2019propagation}.
Generically, kirigami perforated on non-developable surfaces (e.g., spherical surface) can benefit the design of shape-adaptive devices such as wearable sensors \cite{Yang2018Kirigami} and curvy imagers \cite{ko2008hemispherical, rao2021curvy}.
But relevant research is still absent.
Therefore, it is of great significance to develop general theories on the deployments and energy landscapes of kirigami structures covering freeform cutting patterns and Euclidean and non-Euclidean design spaces.

\begin{figure}[!tp]
	\centering
	\includegraphics[width=8.6 cm]{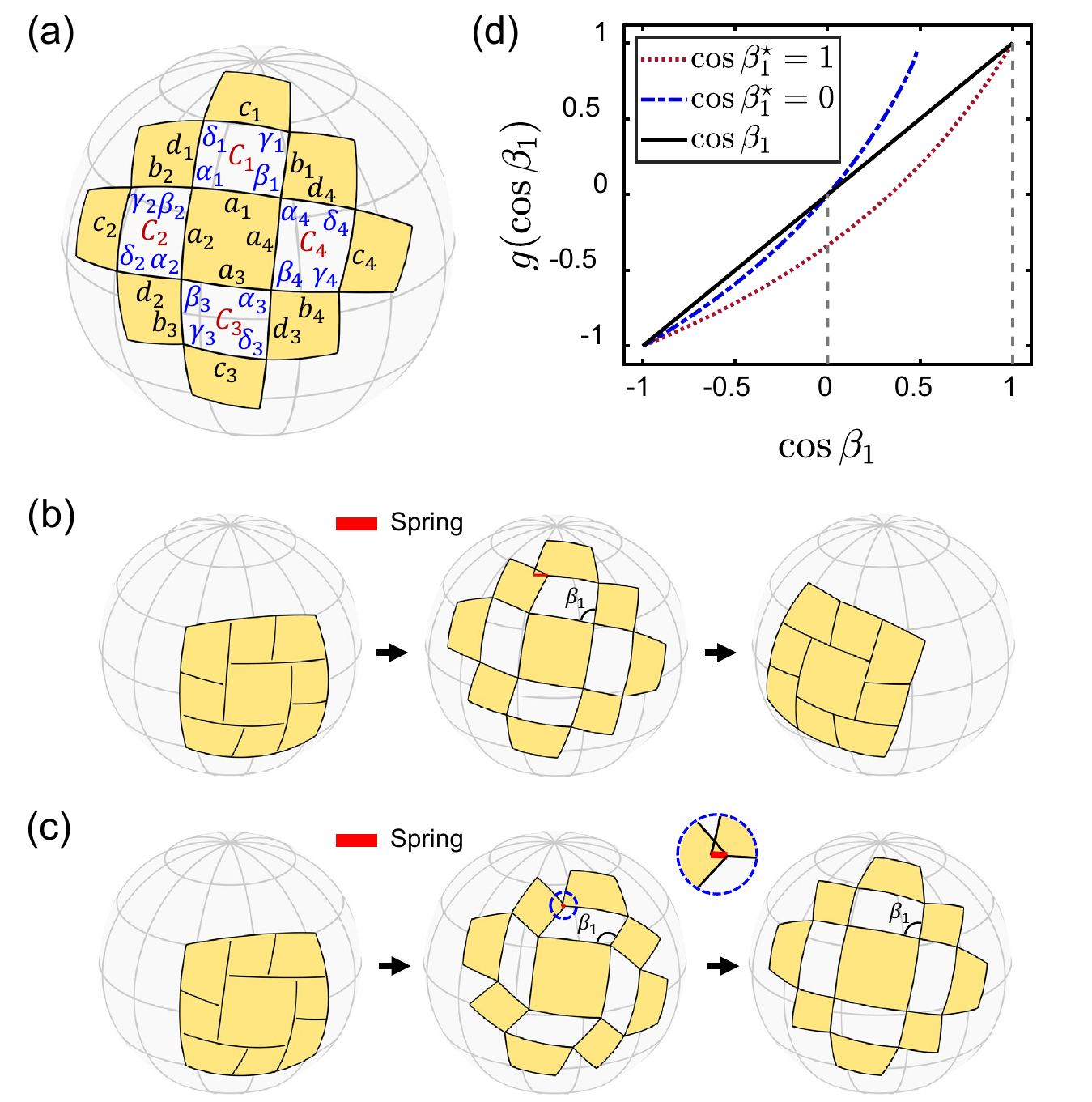}
	\caption {(a) The geometric notations of a deployed $3\times3$ SQK tessellation. (b) A $3\times3$ SQK tessellation at $\cos\beta_1=-1$ (left), 0 (middle), and 1 (right).  { We replace the hinge at the {  top-central} vertex with a spring and permit the overlap of the rigid panels, such that the incompatible configuration (middle) is determined by minimizing the spring elongation.} (c) The optimized $3\times3$ SQK tessellation at $\cos\beta_1=-1$ (left), $-0.5$ (middle), and 0 (right). The compatible configuration is shifted from $\cos\beta_1=1$ to $0$.  (d) Curves of the loop functions.}
	\label{fig:comp}
\end{figure}

In this Letter, we focus on spherical quadrilateral kirigami (SQK) tessellations---the geodesic cuts divide curved sheets into $M$ columns and $N$ rows of arrayed quadrilateral panels [Fig.~\ref{fig:intro} { (see Supplemental Material: Sec.~IX \cite{supp} for fabrication details)}] that are (ideally) connected by free rotational hinges at the corner \footnote{We focus on patterns with a {\it compact} configuration. There are also kirigami patterns with no compact states, for which the cuts are voids of actual area, such as the lattice kirigami \cite{Castle2014Making} and rhombi-slit kirigami \cite{zheng2021continuum}.}---and prove the following compatibility theorem:
\begin{theorem}
	An SQK tessellation has either one or two compatible configurations.
	\label{th:spmn}
\end{theorem}
First, we demonstrate the validity of this theorem for basic $3\times3$ SQK tessellations [Fig.~\ref{fig:comp}(a)] by investigating the corresponding {\it compatibility condition} \footnote{In our discussion, the term \emph{tessellation} stands for $M\times N$ kirigami patterns with $M\geq3$ and $N\geq3$; we refer to patterns with $M=2$ or $N=2$ as \emph{kirigami strip}, which are always rigidly deployable (see Fig.~S1 and Supplemental Material: Sec.~I \cite {supp}).}.
Each solution of the compatibility condition stands for a unique {\it compatible configuration}.
We will verify that the number of such solutions is either one or two, depending on the geometry of the given kirigami pattern.
Then, the proof is accomplished by the fact that the compatibility of an $M\times N$ tessellation requires any of its $3\times3$ parts to be compatible.
Further, we give a corollary for implementing a spherical tessellation which guarantees two compatible configurations, and another one for a floppy planar tessellation with infinite compatible states.
Finally, we design a reconfigurable kirigami structure with a dome-like deployed compatible configuration.

\emph{Compatibility theorem.}
We start by investigating $3\times3$ SQK tessellations built on a sphere with Gaussian curvature $K$.
Based on spherical trigonometry \cite{todhunter1863spherical}, we restrict the slits to be minor arcs of great circles, and employ notations in Fig.~\ref{fig:comp}(a) to formulate the compatibility condition.
One can observe that, first, there are four { connected} quadrilateral slits $C_i$ surrounding the inner panel, with side lengths $a_i$, $b_i$, $c_i$, $d_i \in (0,\pi/\sqrt{K})$, and opening angles $\alpha_i$, $\beta_i$, $\gamma_i$, $\delta_i \in (0,\pi)$ for $i=1,2,3,4$.
Second, since the slits are geodesic lines at the undeployed state, the side lengths satisfy $a_i+b_i=c_i+d_i$.
Given the side lengths of a slit $C_i$, the opening angles $\beta_i$, $\gamma_i$, and $\delta_i$ are uniquely determined as functions of $\alpha_i$ (see Supplemental Material: Sec.~II \cite{supp}).
As a result, the shape of a slit relies on its neighbor by the conserved relations $\alpha_{i}=\pi-\beta_{i+1}$, for $i=1,2,3,4$ ({ $\beta_5$ is defined as $\beta_1$, and the same cyclic relationship follows below for other quantities}).
That is, the opening angles in adjacent slits (say, $\beta_i$ and $\beta_{i+1}$) are in a one-to-one correspondence, which we denote by $\cos\beta_{i}=g_i(\cos\beta_{i+1})$. The explicit expressions of $g_i$ are provided in Supplemental Material: Sec.~II \cite{supp}.
Then we define the loop function $g \triangleq g_1\circ g_2\circ g_3\circ g_4$ for $(g_i\circ g_{i+1})(\cdot)=g_i[g_{i+1}(\cdot)]$.
From the expressions of $g_i$, we observe that $g$ is smooth on the feasible domain $ \cos\beta_{1}\in[-1,c_r]$ for $c_r\in(-1,1]$.
The upper bound $c_r$ represents the stage where at least one of opening angles $\alpha_i$ and $\gamma_i$ reaches $\pi$ upon deployment, and the lower bound $-1$ corresponds to the undeployed state.
A valid compatible configuration requires that the value of $\cos\beta_1$ is preserved around a loop of operations by $g_i$, so that the compatibility condition reads
\begin{equation}
	g(\cos\beta_1) - \cos\beta_1 = 0.
	\label{eq:sp-comp}
\end{equation}
Each root of Eq.~(\ref{eq:sp-comp}) represents one compatible state of a $3\times3$ SQK tessellation.
A trivial solution is $\cos\beta_i=-1$ at the undeployed state.
To explore other compatible configurations, we will next show that $g$ is a strict convex function on $[-1,c_r]$.
The first derivative of $g$ with respect to $\cos\beta_{1}$ is given by $g'(\cos\beta_1) = \prod_{i=1}^4 g'_i(\cos\beta_{i+1})$, in which $g'_i$ can be explicitly expressed as (see Supplemental Material: Sec.~II \cite{supp})
\begin{equation}
	g'_i(\cos\beta_{i+1}) = \frac{\sin (d_i\sqrt{K}) \sin \beta_i \sin \delta_i}{\sin (b_i\sqrt{K}) \sin \alpha_i \sin \gamma_i},
	\label{eq:sp-dfi}
\end{equation}
where we have $\alpha_i=\pi-\beta_{i+1}$, and $\beta_i$, $\gamma_i$, $\delta_i$ are functions of $\alpha_i$ under given side lengths $a_i$, $b_i$, $c_i$, and $d_i$.
Checking the right-hand side of Eq.~(\ref{eq:sp-dfi}), we find $g'_i>0$ on $(-1,c_r)$.
Also, we can prove $g''_i>0$ under the condition $a_i+b_i=c_i+d_i$ (see Supplemental Material: Sec.~III \cite{supp}).
It then follows that $g''>0$ on $(-1,c_r)$.
Adding the smoothness of $g$, we conclude that $g$ is a strict convex function on $[-1,c_r]$.
As a result, Eq.~(\ref{eq:sp-comp}) has at most two roots, and equivalently, a $3\times3$ SQK tessellation has at most two compatible configurations.
It further follows that an $M\times N$ tessellation has at most two compatible states as well, because the number of its compatible configurations cannot exceed that of any of its $3\times3$ parts.
Therefore, Theorem 1 is proved.

\emph{Examples and verification.}
Theorem~\ref{th:spmn} asserts that an SQK tessellation can only have up to one compatible configuration away from its undeployed state. Here we demonstrate a special class of SQK tessellations that are assured to have the deployed compatible configuration.
Assuming $a_i=c_i$ and $b_i=d_i$ for $i=1,2,3,4$, the explicit expressions of the loop function---denoted by $g^e$ in this case---can be derived as (see Supplemental Material: Sec.~III \cite{supp})
\begin{equation}
	\label{eq:sp-para-loop}
	g^e(x) = \frac{(P+Q) x + (P-Q)}{(P+Q) + (P-Q) x},
\end{equation}
in which $P,Q=\prod_{i=1}^4\cos^2[(a_i\pm b_i)\sqrt{K}/2]$.
Then, we solve Eq.~(\ref{eq:sp-comp}) and obtain $\cos\beta_1=\pm 1$, indicating two compact compatible configurations [as shown in Fig.~\ref{fig:comp}(b)].
{ Conversely}, if a $3\times3$ SQK tessellation is compatible at two compact states, we can conclude $a_i=c_i$ and $b_i=d_i$, following the two conditions $a_i+d_i=b_i+c_i$ at $\cos\beta_i=1$ and $a_i+b_i=c_i+d_i$ at $\cos\beta_i=-1$.
In general, for an $M\times N$ SQK tessellation, we can investigate all of its $3\times3$ parts and obtain:
\begin{corollary}
	An SQK tessellation is compatible at two compact configurations if and only if the opposite side lengths are equal for each slit.
	\label{co:sp-para}
\end{corollary}
This corollary is instructive to design SQK patterns with specified compatible states.
For a $3\times3$ SQK tessellation, we define $k^{b}_i$ and $k^{d}_i$ as the ratios by which the slit $C_{i}$ is divided by the intersecting slits $C_{i-1}$ and $C_{i+1}$, i.e., $k^{b}_i=b_{i}/(a_{i}+b_{i})$ and $k^{d}_i=d_{i}/(c_{i}+d_{i})$.
Fixing the boundary vertices, the undeloyed kirigami is uniquely determined by $k^{b}_i$ and $k^{d}_i$ via solving a non-linear equation system, and the deployed state is then given by applying the deformation induced from the reference opening angle $\beta_1$.
If we assign the cutting ratios with $k^{b}_i=k^{d}_i$, the tessellation will be compatible at two compact configurations.
We can further optimize $k^{b}_i$ and $k^{d}_i$ to shift the second compatible configuration from $\cos\beta_1^\star=1$ to $\cos\beta_1^\star\in(-1,1)$.
These formulations are provided in Supplemental Material: Sec.~IV \cite{supp}.

A $3\times3$ SQK tessellation with $k^{b}_i = k^{d}_i = 0.4$ is illustrated in Fig.~\ref{fig:comp}(b).
The slits are prescribed on a spherical square of side length $\pi/3$ and Gaussian curvature $K=1$.
This tessellation has two compact states at $\cos\beta_1^\star=\pm 1$ [Fig.~\ref{fig:comp}(b), left and right], and is incompatible at $\cos\beta_1\in(-1,1)$ [Fig.~\ref{fig:comp}(b), middle].
An optimized tessellation compatible at $\cos\beta_1^\star=0$ is illustrated in Fig.~\ref{fig:comp}(c).
One can observe that the kirigami pattern only changes slightly after optimization [i.e., $k^{b}_i \approx k^{d}_i \approx 0.4$, as shown in Fig.~\ref{fig:comp} (c), left], while the compatible state is converted from $\cos\beta_1^\star=1$ to $0$ [Fig.~\ref{fig:comp}(c), right].
This high sensitivity to the small changes of the reference pattern arises from the high non-linearity of the loop function $g$ with respect to $a_i$, $b_i$, $c_i$, and $d_i$.
The plots of loop functions for these two tessellations are shown in Fig.~\ref{fig:comp}(d).
We can see that both curves are convex and intersect with the $45^\circ$ rising line twice, indicating that there exist two compatible configurations.
For comparison, the $3\times3$ SQK tessellations with a single compatible state are shown in Fig.~S3.

\textit{Rigid-to-floppy transition.} If the Gaussian curvature $K$ is sufficiently small, we can expect that the SQK tessellations are approximately located on a plane, degenerating into planar quadrilateral kirigami (PQK) tessellations.
When $K=0$, Eq.~(\ref{eq:sp-para-loop}) becomes $g^e(x)=x$, so that the compatibility condition always holds for $3\times3$ PQK tessellations with $a_i=c_i$ and $b_i=d_i$ (i.e., the slits form parallelograms).
Otherwise, if $a_i\neq c_i$ or $b_i\neq d_i$, we can verify that the degenerate $g_i'$ and $g_i''$ are always positive (see Supplemental Material: Sec.~V. \cite{supp}).
Hence, Theorem \ref{th:spmn} still holds under this circumstance.
Generally, we can investigate all the $3\times3$ parts of $M\times N$ PQK tessellations and obtain:
\begin{corollary}
	A PQK tessellation is rigidly deployable between two compact configurations if and only if all the slits form parallelograms. Otherwise, a PQK tessellation has either one or two compatible configurations.
	\label{co:pl-para}
\end{corollary}
In the inspiring work \cite{choi2021compact}, Choi {\it et al}. proved that a planar kirigami tessellation with kite-shape slits is rigidly deployable if and only if all the slits are rhombuses.
Corollary \ref{co:pl-para} further extends the design space of quadrilateral kirigami, as a floppy mechanism, to a much broader domain, i.e., from rhombus to parallelogram slits.
	
Corollaries \ref{co:sp-para} and \ref{co:pl-para} are obtained from the geometric compatibility. Actually, they reflect the physical insights on the rigid-to-floppy transition from curved to flat kirigami.
We now examine the connection between Gaussian curvature and the rigidity of SQK tessellations.
To this end, we develop a {\it single-spring model}---the kirigami is represented by a system of hinge-connected rigid panels, except that one hinge is replaced by a linear spring with stiffness $k_S$, as illustrated in Fig.~\ref{fig:comp}(b). In this way, the compatible configuration corresponds to a zero elongation of the spring, whereas an incompatible configuration corresponds to a non-zero elongation.
Then, the incompatibility or rigidity of the system can be characterized by the elastic energy $E_S=(k_S/2)\Delta_S^2$, where $\Delta_S$ is the elongation of the spring.
We use Taylor's series to expand the scaled energy $E_S/(k_Sb_1^2)$ at $K=0$ (see Supplemental Material: Sec.~VI.A \cite{supp}):
\begin{equation}
	\frac{E_S}{k_Sb_1^2} = \frac{1}{8}\left[\sum_{i=1}^4(a_ib_i)\right]^2K^2\sin^4\beta_1 + O[L^6K^3],
	\label{eq:esSeries}
\end{equation}
where $L=\max\{a_1,b_1,...,a_4,b_4\}$, and $L^2K\ll1$.
The leading term in Eq.~(\ref{eq:esSeries}) reflects the competing roles of the spherical surface area ($\sim 1/K$) and the slit size ($\sim \sum_{i=1}^4 a_{i}b_{i} \sim L^2$).
It clearly and explicitly shows that, when $K=0$, the energy is zero for any value of $\beta_1$, which means that there is an entire path of zero-energy deployments on a plane; however, when
$K\neq 0$, there are only two isolated zero-energy states: $\beta_1=\pi$ (undeployed) and $\beta_1=0$ (deployed and compact). Moreover,  Eq.~(\ref{eq:esSeries}) can also be related to the loop function $g^e$ as follows (see Supplemental Material: Sec.~VI.A \cite{supp}):
\begin{equation}\label{supl58}
	\frac{E_S}{k_Sb_1^2} = \frac{1}{2} [g^e(\cos\beta_1) - \cos\beta_1]^2 + O[L^6K^3],
\end{equation}
which indicates that the degree of incompatibility $|g^e(\cos\beta_1) - \cos\beta_1|$ characterizes the magnitude of the elastic energy. Since $g^e(\cos\beta_1) - \cos\beta_1=0$ only
has two roots [Fig.~\ref{fig:comp}(d)], Eq.~(\ref{supl58}) also reveals two zero-energy configurations.
{ Note that the apparent rigidity of the two-configuration result will vanish if we introduce additional freedom to move in the radial direction (see Supplemental Material: Sec.~VIII. \cite{supp}).}

\begin{figure}[!t]
\centering
\includegraphics[width=8.6 cm]{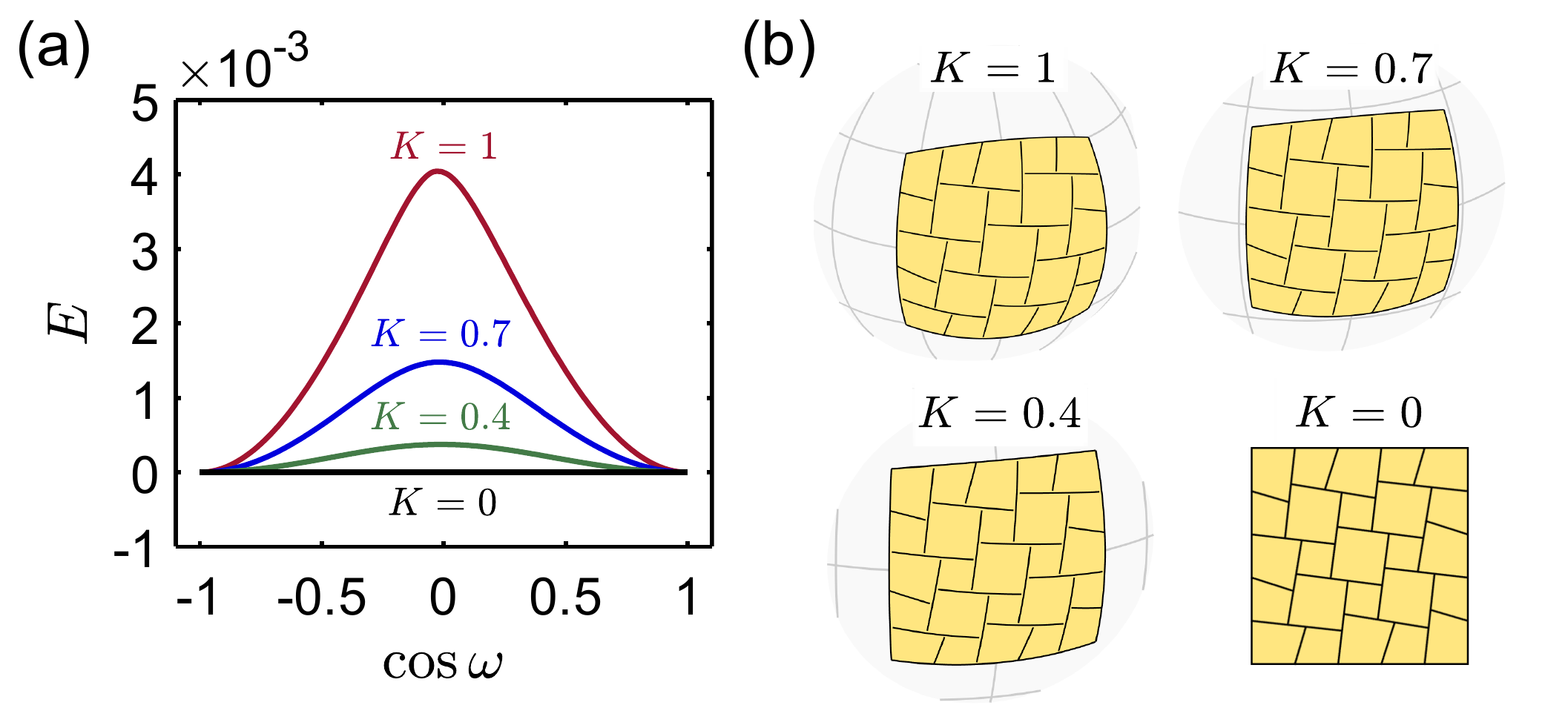}
\caption{ (a) The evolutions of elastic energy $E$ of $5\times5$ SQK and PQK shells with equal opposite side lengths of slits for different values of Gaussian curvature $K=1$, $0.7$, $0.4$, and $0$. The energy curve of a PQK tessellation ($K=0$) is constantly zero. (b) Undeployed patterns of the tessellations. The aspect ratios of slits are fixed as $0.4$.}
\label{fig:energy}
\end{figure}

The evolution of the energy upon deployment can be simulated by a {\it multispring model}, in which the kirigami is represented by hinge-connected springs along the edges and diagonals of panels.
In this model, the elastic energy of the deployed tessellations can be written as $E\left(\bf{Y}\right) = \sum\nolimits_{n}k_n[l_n\left(\bf{Y}\right)-l^0_n]^2/2$, where $\bf{Y}$ is the array of panel-vertex positions, $l_n$ the spring length numbered by the index $n$, $l^0_n$ the rest length at the undeformed state, and $k_n$ the spring stiffness (set to be $1/l^0_n$).
The deployed configurations are determined by incrementally increasing the kinematic parameter $\cos\omega$---defined by $\cos\beta_1$ of the lower-left $3\times3$ tessellation [Fig.~\ref{fig:intro}(b)]---from the undeployed state.
At each step, we minimize the energy $E\left(\bf{Y}\right)$ taking positions of all the vertices as variables, which are constrained to form an SQK pattern, and enforce the opening angle $\omega$ (see Supplemental Material: Sec.~VI.B \cite{supp}).
Figure~\ref{fig:energy}(a) demonstrates the energy curves of SQK shells perforated on a spherical square of fixed side length $s=\pi/3$ for different Gaussian curvatures. While decreasing $K$ from 1 to 0.4 [Fig.~\ref{fig:energy}(b)], the energy barrier drops significantly.
If $K=0$, the SQK tessellation degenerates to a PQK tessellation, which is rigidly deployable (floppy) with zero energy of deformations.
We illustrate the deployed configurations of these kirigami tessellations in Fig.~S6.
{ We additionally show the energy curves with deployed compatible states at $\cos\omega\in(-1,1)$ in Fig.~S7.}

\begin{figure}[!t]
	\centering
	\includegraphics[width=8.6 cm]{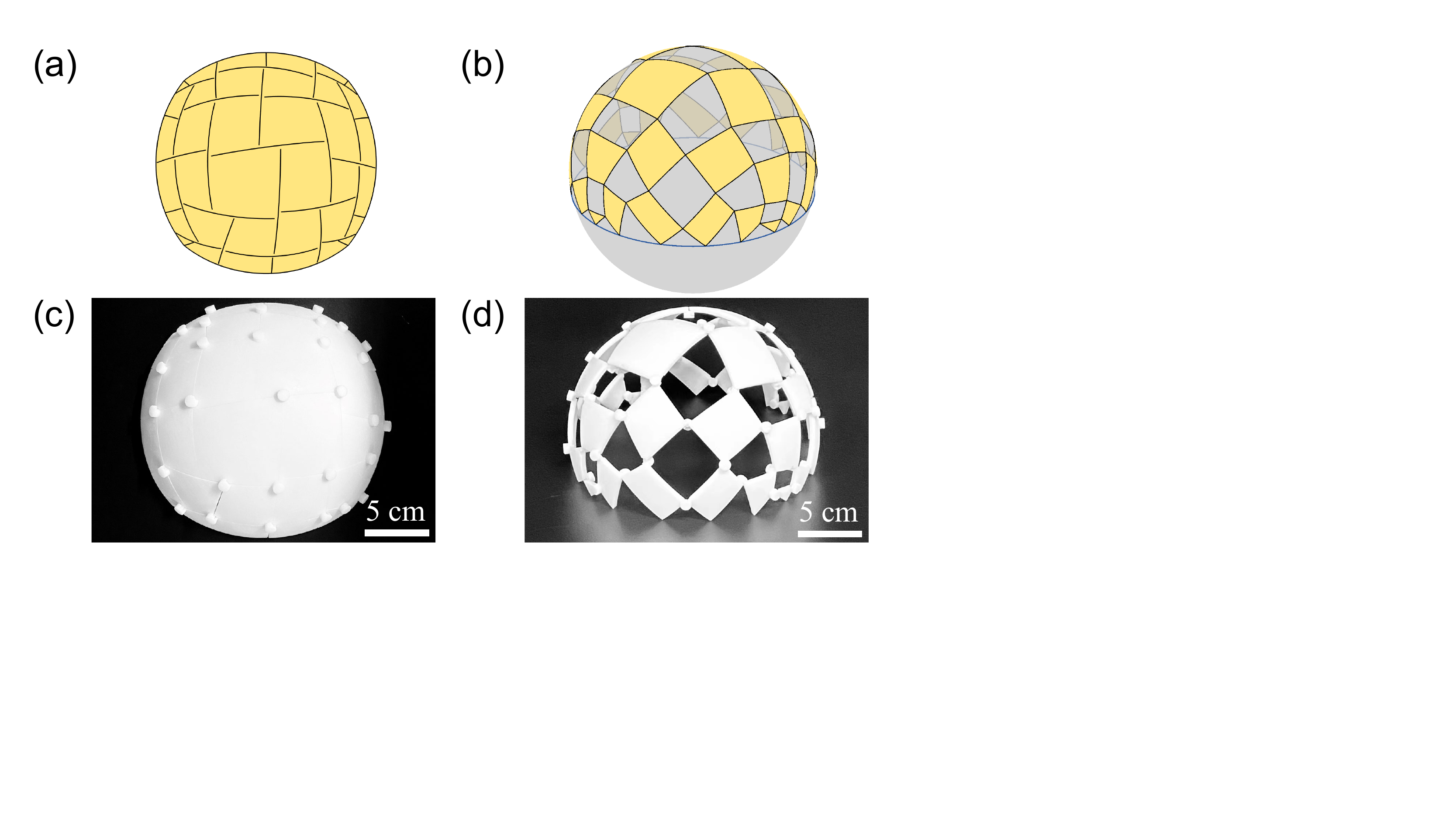}
	\caption{Reconfigurable SQK dome-like shell structure. (a) The kirigami pattern is perforated on a spherical square of side length $s=0.465\pi$ and Gaussian curvature $K=1$. (b) The deployed configuration covers a spherical dome of height $h=1.2$. { (c) and (d) The physical model made of resin.}}
	\label{fig:dome}
\end{figure}

{\it Shape-morphing structures.} As a final demonstration for potential applications of our theorem, we address how to design the cutting patterns to achieve compatible deployed configurations with desired shapes.
We start from a $6\times6$ square SQK shell structure of side length $s=0.465\pi$ and Gaussian curvature $K=1$, and then optimize the locations of vertices via minimizing the distance between the deployed outer vertices and the boundary of a spherical dome of radius $r=1$ and height $h=1.2$.
Details of the optimization framework are given in Supplemental Material: Sec.~VII \cite{supp}.
Figure~\ref{fig:dome} demonstrates the shell structure { (see Supplemental Material: Sec.~IX \cite{supp} for fabrication details)}.
The covering area of the undeployed pattern and the deployed dome can be calculated by $\tan(S_{\rm squa.}/4) = \sin^2(s/2)\sqrt{\sec s}$ and $S_{\rm dome} = 2\pi r h$, respectively.
Thus, the expansion ratio of the area is $S_{\rm dome}/S_{\rm squa.}\approx2.0$.
According to Theorem~\ref{th:spmn}, the deployed configuration is at an isolated compatible state, which is ideally rigid, so that it can form a stable structure for potential applications such as tents and roofs { (see Movie 1 in Supplemental Material \cite{supp})}.
Moreover, shape-morphing mechanisms can be realized by PQK tessellations { with different topologies \cite{dang2021theorem}.}

{\it Conclusion and discussion.} In summary, we show that spherical quadrilateral kirigami tessellations can only be compatible at isolated configurations, whereas planar kirigami quadrilateral tessellations with parallelogram slits have infinite and continuous compatible states.
We develop single-spring and multispring models to explicitly analyze and simulate the evolution of energy along the deployment paths for different values of Gaussian curvature, which characterizes the rigid-to-floppy transition.
Since the deformation energy drops appreciably near the flat surface, the slightly curved kirigami shells are expected to be a promising candidate of pseudo-mechanisms, as those presented in Refs.~\cite{Singh2021design, zheng2021continuum}.

{ The compatibility theorem and its corollaries reveal the role of curvature in determining the deployment behaviors of kirigami systems.}
The effect of curvature on constrained morphology of stable structures can also be observed in various physical phenomena in nature such as the growth of nanoshells \cite{Bao2011One} and rigid colloidal crystals \cite{meng2014elastic} on spherical substrates.
More curvature-induced scenarios, such as the buckling of non-Euclidean kirigami shells---the counterpart of buckling-induced planar kirigami \cite{rafsanjani2017Buckling}---can be further investigated.
Finally, the compatibility in our theorem relies purely on the size-independent geometry of the prescribed slits, so that the applicability is rooted in diverse materials and various scales.

\begin{acknowledgments}
	X.D., H.D. and J.W. thank the National Natural Science Foundation of China (Grant Nos.~11991033, 91848201, and 11521202) for support of this work.
	The authors thank Xiying Li and Yu Zou for assistance in 3D printing. We also thank Lu Lu and Paul Plucinsky for helpful discussions.
\end{acknowledgments}

\bibliography{kirigami}

\end{document}


\title{Supplemental Material: \\Theorem on the Compatibility of Spherical Kirigami Tessellations}

\author{Xiangxin Dang}
\affiliation{State Key Laboratory for Turbulence and Complex Systems, Department of Mechanics and Engineering Science, College of Engineering, Peking University, Beijing 100871, China}
\author{Fan Feng}
\affiliation{Cavendish Laboratory, University of Cambridge, Cambridge CB3 0HE, UK}
\author{Huiling Duan}
\author{Jianxiang Wang}
\email{jxwang@pku.edu.cn}
\affiliation{State Key Laboratory for Turbulence and Complex Systems, Department of Mechanics and Engineering Science, College of Engineering, Peking University, Beijing 100871, China}
\affiliation{CAPT-HEDPS, and IFSA Collaborative Innovation Center of MoE, College of Engineering, Peking University, Beijing 100871, China}

\date{\today}

\maketitle


\setcounter{equation}{0}
\setcounter{figure}{0}
\setcounter{table}{0}
\setcounter{page}{1}
\setcounter{section}{0}
\makeatletter
\renewcommand{\theequation}{S\arabic{equation}}
\renewcommand{\thefigure}{S\arabic{figure}}
\renewcommand{\bibnumfmt}[1]{[S#1]}
\renewcommand{\citenumfont}[1]{S#1}

\section{Kirigami strips}

\begin{figure}[!h]
	\centering
	\includegraphics[width=0.8\columnwidth]{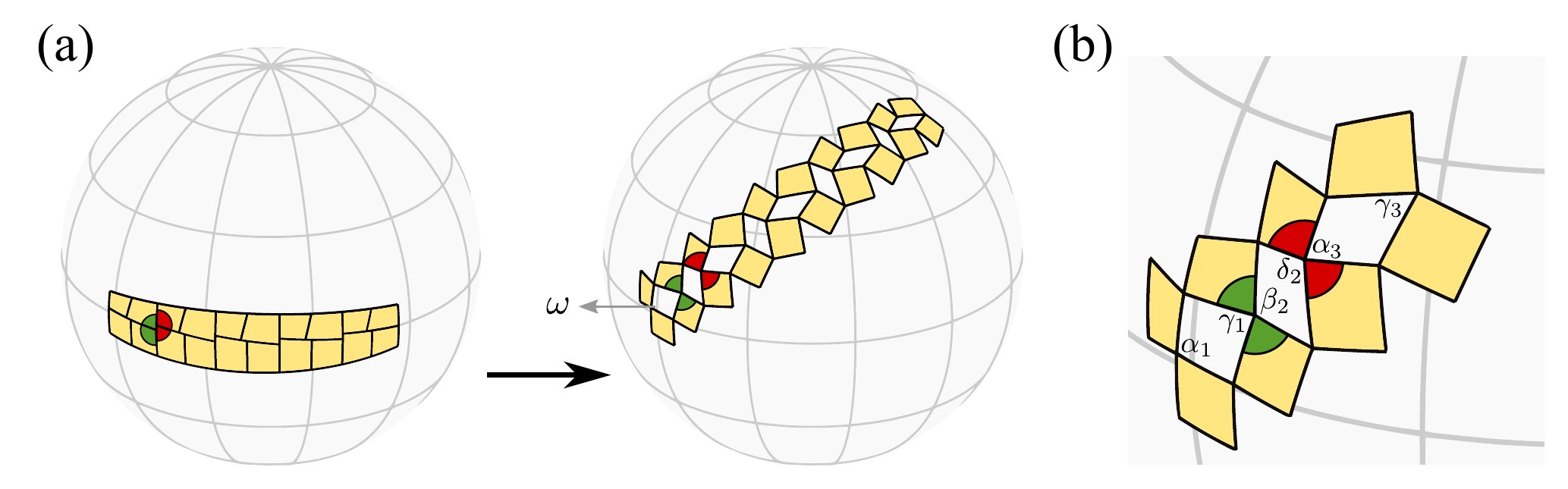}
	\caption{SQK strips. (a) A SQK strip and its deployed configuration at $\omega=0.5\pi$. (b) The opening angles can be transmitted over the slits by the constraints that the opposite angles of adjacent slits sum to $\pi$.}
	\label{fig:SuppSPStrip}
\end{figure}

We refer to SQK with more than two columns and rows as SQK tessellations, and those with only two columns or rows as SQK strips. 
A $10\times2$ SQK strip is illustrated in Fig.~\ref{fig:SuppSPStrip}(a). 
The values of opening angles can be transmitted throughout the kirigami strip by the fact that each single quadrilateral void has one free opening angle and the opposite angles at the same vertex sum to $\pi$. 
For example, as shown in Fig.~\ref{fig:SuppSPStrip}(b), we can see that $\gamma_1+\beta_2=\pi$ and $\delta_2+\alpha_3=\pi$. 
As a result, the kirigami strip can be continuously deployed with a single degree of freedom.
The deformation formulation of the SQK strips can be included in the framework of general $M\times N$ SQK tessellations, which we will show below.

\section{Geometry of a single slit}

\begin{figure}[!t]
	\centering
	\includegraphics[width=0.35\textwidth]{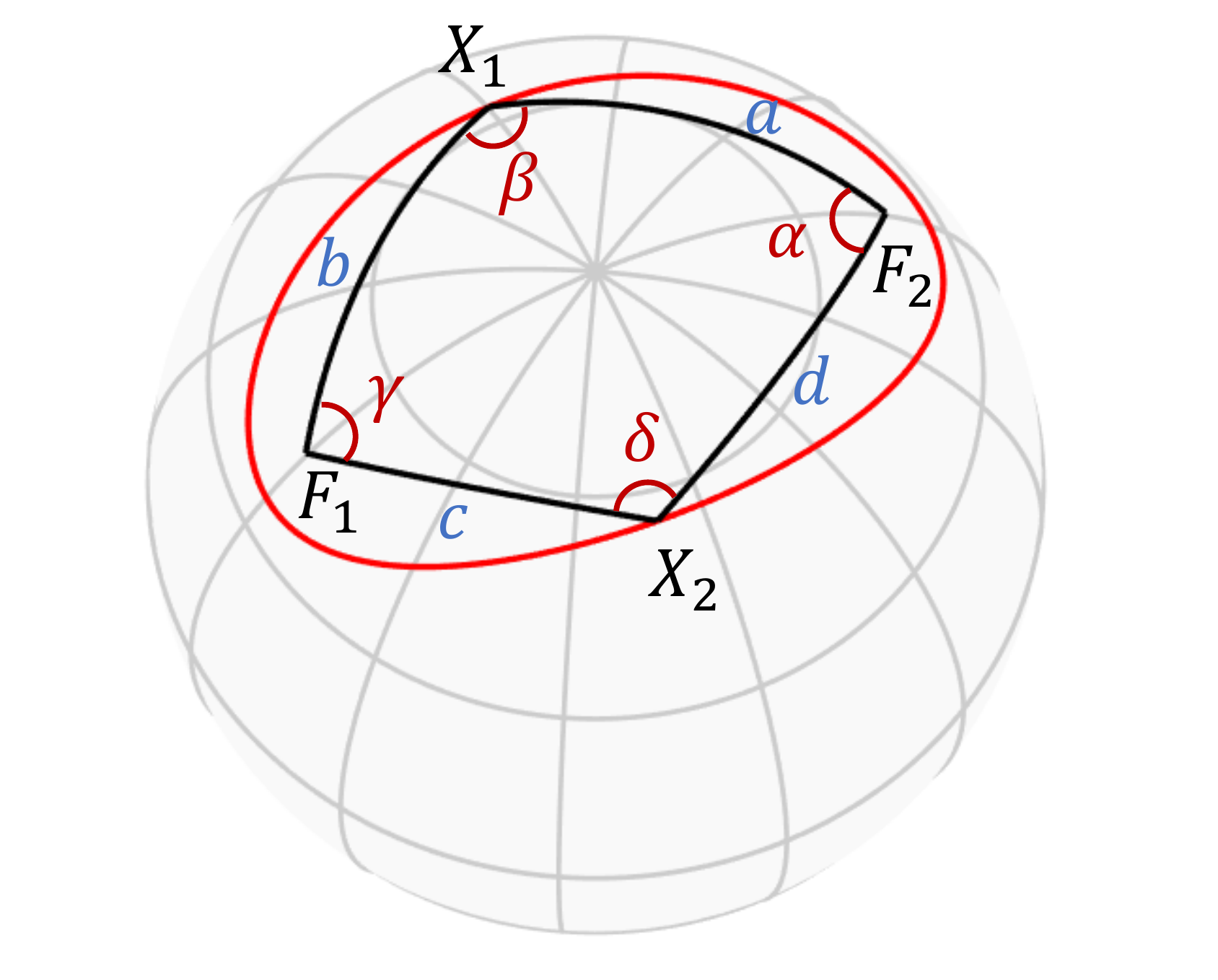}
	\caption{Notations of a convex spherical quadrilateral void $F_2 X_1 F_1 X_2$ of side lengths $a,~b,~c,~d$, and opening angles $\alpha,~\beta,~\gamma,~\delta$. This void is formed by a deployed slit of SQK strips or tessellations, so that the side lengths satisfy $a+b=c+d$. In addition, the quadrilateral void can be determined by two points $X_1$, $X_2$ on a spherical ellipse with foci $F_1$, $F_2$. }
	\label{fig:SuppSPEllipse}
\end{figure}

A single slit of SQK strips or tessellations is deployed to a convex spherical quadrilateral void. 
We use $K$ to denote the Gaussian curvature of the sphere.
As shown in Fig.~\ref{fig:SuppSPEllipse}, the spherical quadrilateral $F_2 X_1 F_1 X_2$ is formed by the great-circle arcs $a,b,c,d\in(0,\pi/\sqrt{K})$ with opening angles $\alpha,\beta,\gamma,\delta\in(0,\pi)$. 
The quadrilateral void can be divided into two spherical triangles ${F_2 X_1 X_2}$ and ${X_1 F_1 X_2}$ by connecting the vertices $X_1$ and $X_2$ via a geodesic curve, or into spherical triangles ${F_2 X_1 F_1}$ and ${F_2 F_1 X_2}$ by connecting the vertices $F_1$ and $F_2$.
Following the spherical cosine rules, we have
\begin{align}
	\label{eq:supp-sp-cos-1}
	&\cos{(a\sqrt{K})}\cos{(d\sqrt{K})}+\sin{(a\sqrt{K})}\sin{(d\sqrt{K})}\cos{\alpha} =  \cos{(b\sqrt{K})}\cos{(c\sqrt{K})}+\sin{(b\sqrt{K})}\sin{(c\sqrt{K})}\cos{\gamma},\\
	\label{eq:supp-sp-cos-2}
	&\cos{(a\sqrt{K})}\cos{(b\sqrt{K})}+\sin{(a\sqrt{K})}\sin{(b\sqrt{K})}\cos{\beta} =  \cos{(c\sqrt{K})}\cos{(d\sqrt{K})}+\sin{(c\sqrt{K})}\sin{(d\sqrt{K})}\cos{\delta}.
\end{align}
The differential forms of Eqs.~(\ref{eq:supp-sp-cos-1}) and (\ref{eq:supp-sp-cos-2}) are
\begin{eqnarray}
	\label{eq:supp-sp-dAlpha-dGamma}
	\frac{\dif\gamma}{\dif\alpha} &=& \frac{\sin{(a\sqrt{K})}\sin{(d\sqrt{K})}\sin{\alpha}}{\sin{(b\sqrt{K})}\sin{(c\sqrt{K})}\sin{\gamma}},\\
	\label{eq:supp-sp-dBeta-dDelta}
	\frac{\dif\delta}{\dif\beta} &=& \frac{\sin{(a\sqrt{K})}\sin{(b\sqrt{K})}\sin{\beta}}{\sin{(c\sqrt{K})}\sin{(d\sqrt{K})}\sin{\delta}}.
\end{eqnarray}
Denote the area of the spherical triangle ${F_2 X_1 X_2}$ as $S_\alpha$. 
The area $S_\alpha$ can be determined by the side lengths $a,d$, and the included angle $\alpha$:
\begin{equation}
	\tan(K S_\alpha/2)=\frac{\tan(a\sqrt{K}/2)\tan(d\sqrt{K}/2)\sin{\alpha}}{1+\tan(a\sqrt{K}/2)\tan(d\sqrt{K}/2)\cos{\alpha}}.
	\label{eq:supp-tanSalpha2}
\end{equation}
We will find the following trigonometric identities useful:
\begin{eqnarray}
	\label{eq:supp-sp-sinSAlpha}
	\sin (K S_{\alpha})=\frac{2 \tan (K S_{\alpha}/2)} {1+\tan ^2(K S_{\alpha}/2)},\\
	\label{eq:supp-sp-cosSAlpha}
	\cos (K S_{\alpha})=\frac{1-\tan ^2(K S_{\alpha}/2)}{1+\tan ^2(K S_{\alpha}/2)},\\
	\label{eq:supp-sp-tanSAlpha}
	\tan (K S_{\alpha})=\frac{2 \tan (K S_{\alpha}/2)} {1-\tan ^2(K S_{\alpha}/2)}.
\end{eqnarray}
We also have the following differential relationship:
\begin{eqnarray}
	\label{eq:supp-sp-dSAlpha}
	\frac{\dif (K S_\alpha)}{\dif\alpha}=\cos^2{(K S_\alpha)}\frac{\dif \tan (K S_\alpha)}{\dif\alpha}.
\end{eqnarray}	
Substituting Eqs.~(\ref{eq:supp-tanSalpha2}), (\ref{eq:supp-sp-cosSAlpha}), (\ref{eq:supp-sp-tanSAlpha}) into Eq.(\ref{eq:supp-sp-dSAlpha}) we obtain
\begin{eqnarray}
	A(\alpha,K) \triangleq 1-\frac{\dif (K S_\alpha)}{\dif\alpha}=\frac{\cos (a\sqrt{K})+\cos (d\sqrt{K})}{\sin (a\sqrt{K}) \sin (d\sqrt{K}) \cos \alpha+\cos (a\sqrt{K}) \cos (d\sqrt{K}) +1}.
	\label{eq:supp-sp-dS_alpha}
\end{eqnarray}
In the same way, we denote $S_{\gamma}$ as the area of the spherical triangle $X_1F_1X_2$, and define the function $B(\gamma,K)$ by 
\begin{eqnarray}
	B(\gamma,K) \triangleq 1-\frac{\dif (K S_\gamma)}{\dif\gamma}=\frac{\cos (b\sqrt{K}) +\cos (c\sqrt{K}) }{\sin (b\sqrt{K}) \sin (c\sqrt{K}) \cos \gamma +\cos (b\sqrt{K}) \cos (c\sqrt{K}) +1}.
	\label{eq:supp-sp-dS_gamma}
\end{eqnarray}
The interior angle summation and the spherical quadrilateral area have the following relationship:
\begin{eqnarray}
	\alpha+\beta+\gamma+\delta = 2\pi+K S_\alpha+K S_\gamma.
	\label{eq:supp-sp-id}
\end{eqnarray}
The differential form of Eq.~(\ref{eq:supp-sp-id}) is
\begin{equation}
	\dif\alpha+\dif\beta+\dif\gamma+\dif\delta = \dif( K S_\alpha)+\dif (K S_\gamma).
	\label{eq:supp-sp-d-id}
\end{equation}
Eqs.~(\ref{eq:supp-sp-cos-1}), (\ref{eq:supp-sp-cos-2}), (\ref{eq:supp-sp-id}) indicate that the spherical quadrilateral void has a single degree-of-freedom upon deployment (three equations with four unknowns $\alpha$, $\beta$, $\gamma$, $\delta$).
Substituting Eqs.~(\ref{eq:supp-sp-dAlpha-dGamma}), (\ref{eq:supp-sp-dBeta-dDelta}), (\ref{eq:supp-sp-dS_alpha}), (\ref{eq:supp-sp-dS_gamma}) into Eq.~(\ref{eq:supp-sp-d-id}), we obtain
\begin{equation}
	\frac{\dif\beta}{\dif\alpha}
	+\frac{\sin{(a\sqrt{K})}\sin{(b\sqrt{K})}\sin{\beta}}{\sin{(c\sqrt{K})}\sin{(d\sqrt{K})}\sin{\delta}}\frac{\dif\beta}{\dif\alpha}+A(\alpha,K)+B(\gamma,K)\frac{\sin{(a\sqrt{K})}\sin{(d\sqrt{K})}\sin{\alpha}}{\sin{(b\sqrt{K})}\sin{(c\sqrt{K})}\sin{\gamma}}=0.
	\label{eq:supp-sp-dBeta-dAlpha-0}
\end{equation}
Besides, we have the identity
\begin{equation}
	\frac{\sin (c\sqrt{K}) \sin (d\sqrt{K}) \sin \delta+\sin (a\sqrt{K}) \sin (b\sqrt{K}) \sin \beta}{B(\gamma,K)\sin (a\sqrt{K}) \sin (d\sqrt{K}) \sin \alpha+ A(\alpha,K) \sin (b\sqrt{K}) \sin (c\sqrt{K}) \sin \gamma}=1.
	\label{eq:supp-sp-id-2}
\end{equation}
{We will prove Eq.~(\ref{eq:supp-sp-id-2}) later in Section~III.}
Substituting Eq.~(\ref{eq:supp-sp-id-2}) into Eq.~(\ref{eq:supp-sp-dBeta-dAlpha-0}), we obtain
\begin{equation}
	\frac{\dif\beta}{\dif\alpha} = -\frac{\sin (d\sqrt{K}) \sin \delta}{\sin (b\sqrt{K}) \sin \gamma}.
	\label{eq:supp-sp-dBeta-dAlpha}
\end{equation}
We define the relationship between $\alpha$ and $\beta$ by 
\begin{equation}
    \cos \beta = \bar g(\cos\alpha;a,b,c,d,K).
    \label{eq:cb}
\end{equation}
Since $\beta=\sphericalangle F_1X_1X_2+\sphericalangle X_2X_1F_2$, we can use the spherical cosine rules and derive the explicit expression of $\bar g$:
\begin{equation}
	\begin{aligned}
	&\bar g[\cos\alpha;a,b,c,d,K]=
	\cos\left\{	\begin{aligned}
		&\arccos\left[\frac{\cos (c\sqrt{K})-\cos (b\sqrt{K}) \cos (e\sqrt{K})}{\sin (b\sqrt{K}) \sin (e\sqrt{K})}\right]\\
		+&\arccos\left[\frac{\cos (d\sqrt{K})-\cos (a\sqrt{K}) \cos (e\sqrt{K})}{\sin (a\sqrt{K}) \sin (e\sqrt{K})}\right]
	\end{aligned}
	\right\},\\
	&e\sqrt{K}=\arccos[\cos{(a\sqrt{K})}\cos{(d\sqrt{K})}+\sin{(a\sqrt{K})}\sin{(d\sqrt{K})}\cos{\alpha}],
	\end{aligned}
	\label{sp-cos-1}
\end{equation}
where $e$ is the length of the diagonal ${X_1X_2}$.
If the side lengths $a,b,c,d$ are given, the opening angles $\beta$, $\gamma$, $\delta$ are uniquely and explicitly determined by $\alpha$ according to Eqs.~(\ref{eq:supp-sp-cos-1}), (\ref{eq:supp-sp-cos-2}), (\ref{eq:cb}), and (\ref{sp-cos-1}).
The derivative of $\bar g$ with respect to $\cos\alpha$ can be obtained from Eqs.~(\ref{eq:supp-sp-dBeta-dAlpha}) and (\ref{eq:cb}):
\begin{equation}
	\bar g'[\cos\alpha;a,b,c,d,K] = \frac{d \cos\beta}{d \cos\alpha} = -\frac{\sin (d\sqrt{K}) \sin \beta \sin \delta}{\sin (b\sqrt{K}) \sin \alpha \sin \gamma}.
	\label{eq:supp-sp-dg}
\end{equation}
In the Main Text, following the notations in Fig.~2(a), we denote the relationships between adjacent slits in $3\times3$ SQK tessellations by
\begin{equation}
    \cos\beta_{i}=g_i(\cos\beta_{i+1}).
    \label{eq:supp-cosbetai}
\end{equation}
Meanwhile, the opening angle $\beta_{i}$ can also be determined by $\alpha_i$:
\begin{equation}
    \cos\beta_{i}=\bar g[\cos\alpha_{i};a_i,b_i,c_i,d_i,K].
\end{equation}
Remind that we have the following complementary relationships:
\begin{equation}
    \cos\alpha_{i}=-\cos\beta_{i+1}.
    \label{eq:supp-cosalphai}
\end{equation}
Combining Eqs.~(\ref{eq:supp-cosbetai})--(\ref{eq:supp-cosalphai}), we obtain the expressions of $g_i$ and $g'_i$:
\begin{eqnarray}
	g_i(\cos\beta_{i+1}) &=& \bar g[-\cos\beta_{i+1};a_i,b_i,c_i,d_i,K],
	\label{eq:supp-sp-gi}\\
	g'_i(\cos\beta_{i+1}) &=& -\bar g'[-\cos\beta_{i+1};a_i,b_i,c_i,d_i,K],
    \label{eq:supp-sp-dgi}
\end{eqnarray}
where the index $i$ cycles from 1 to 4.

\section{Convexity of the loop function}
Analogous to Eqs.~(\ref{eq:supp-sp-gi}) and (\ref{eq:supp-sp-dgi}), we can write the expressions of $g''_i$:
\begin{eqnarray}
	g''_i(\cos\beta_{i+1}) = \bar g''[-\cos\beta_{i+1};a_i,b_i,c_i,d_i,K].
	\label{eq:supp-sp-ddgi}
\end{eqnarray}
Taking the derivative of $\bar g'$, and combining Eqs.~(\ref{eq:supp-sp-dAlpha-dGamma}), (\ref{eq:supp-sp-dBeta-dDelta}), (\ref{eq:supp-sp-dBeta-dAlpha}), we obtain
\begin{eqnarray}
	\bar g''[\cos\alpha;a,b,c,d,K] =
	\bar k_g
	\left[
	\begin{aligned}
	&-\sin (a\sqrt{K}) \sin (d\sqrt{K}) \sin\alpha\cot \gamma-\sin (b\sqrt{K}) \sin (c\sqrt{K}) \sin\gamma\cot\alpha\\ 
	&-\sin (a\sqrt{K}) \sin (b\sqrt{K}) \sin\beta\cot \delta-\sin (c\sqrt{K}) \sin (d\sqrt{K}) \sin\delta\cot\beta
	\end{aligned}
    \right],
	\label{eq:supp-sp-ddg}
\end{eqnarray}
where the coefficient $\bar k_g$ is
\begin{equation}
\bar k_g=\frac{\sin\beta\sin\delta\sin (d\sqrt{K})}{\sin^2\alpha\sin^2\gamma\sin^2 (b\sqrt{K})\sin (c\sqrt{K})}>0.
\label{eq:kgbar}
\end{equation}
The side lengths $a,b,c,d$ of quadrilateral voids extracted from an SQK strip or tessellation are constrained by $a+b=c+d$, which means that the quadrilateral voids can be determined by two points and two foci of spherical ellipses.
Fig.~\ref{fig:SuppSPEllipse} illustrates a spherical ellipse with major axis length $2l$ and focal distance $2f$, satisfying $0<f\sqrt{K}<l\sqrt{K}<\pi/2$. 
The foci $F_1,F_2$ and two points $X_1,X_2$ on the ellipse determine the spherical quadrilateral $F_2 X_1 F_1 X_2$.
We set up a Cartesian coordinate system located at the spherical center $O=(0,0,0)$, such that the coordinates of the foci $F_1,F_2$ and points $X_1,X_2$ are given by 
\begin{eqnarray}
	\begin{aligned}
        &F_1=\left(-\sin (f\sqrt{K})/\sqrt{K}, 0, \cos (f\sqrt{K})/\sqrt{K}\right),\\
        &F_2=\left(\sin (f\sqrt{K})/\sqrt{K}, 0, \cos (f\sqrt{K})/\sqrt{K}\right),\\
        &X_1=(x_1,y_1,z_1),\\
        &X_2=(x_2,y_2,z_2).
	\end{aligned}
\label{eq:fandx}
\end{eqnarray}
The points $X_1,X_2$ are located at different sides of the plane $OF_1F_2$, and therefore can be parametrized by $\theta_1\in(0,\pi)$ and $\theta_2\in(-\pi,0)$:
\begin{eqnarray}
	\begin{aligned}
	&x_j = \sin (l\sqrt{K})\cos\theta_j/\sqrt{K}, \\
	&y_j = \sqrt{\frac{\sin^2(l\sqrt{K})-\sin^2(f\sqrt{K})}{\cos^2(f\sqrt{K})}}\sin\theta_j/\sqrt{K},\\
	&z_j = \sqrt{\cos^2\theta_j+\frac{\sin^2\theta_j}{\cos^2(f\sqrt{K})}}\cos (l\sqrt{K})/\sqrt{K},
	\end{aligned}
\end{eqnarray}
for $j = 1,2$.
Using the locations of vertices $F_1, F_2, X_1, X_2$, we can obtain the following expressions:
\begin{eqnarray}
	\begin{aligned}
	\sin (d \sqrt{K}) \sin (a\sqrt{K}) \sin \alpha  = K^{5/2}\left[ {\left( {{F_2} \times {X_1}} \right) \times \left( {{F_2} \times {X_2}} \right)} \right] \cdot {F_2},{\rm{ }}\\ 
	\sin (a\sqrt{K}) \sin (b\sqrt{K}) \sin \beta  = K^{5/2}\left[ {\left( {{X_1} \times {F_1}} \right) \times \left( {{X_1} \times {F_2}} \right)} \right] \cdot {X_1},\\
	\sin (b\sqrt{K}) \sin (c\sqrt{K}) \sin \gamma  = K^{5/2}\left[ {\left( {{F_1} \times {X_2}} \right) \times \left( {{F_1} \times {X_1}} \right)} \right] \cdot {F_1},{\rm{ }}\\
	\sin (c\sqrt{K}) \sin (d\sqrt{K}) \sin \delta  = K^{5/2}\left[ {\left( {{X_2} \times {F_2}} \right) \times \left( {{X_2} \times {F_1}} \right)} \right] \cdot {X_2},
	\end{aligned}
\label{eq:sss}
\end{eqnarray}
\begin{eqnarray}
	\begin{aligned}
	\sin (d\sqrt{K}) \sin (a\sqrt{K}) \cos \alpha  = K^2\left( {{F_2} \times {X_1}} \right) \cdot \left( {{F_2} \times {X_2}} \right),{\rm{ }}\\
	\sin (a\sqrt{K}) \sin (b\sqrt{K}) \cos \beta  = K^2\left( {{X_1} \times {F_1}} \right) \cdot \left( {{X_1} \times {F_2}} \right),\\
	\sin (b\sqrt{K}) \sin (c\sqrt{K}) \cos \gamma  = K^2\left( {{F_1} \times {X_2}} \right) \cdot \left( {{F_1} \times {X_1}} \right),{\rm{ }}\\
	\sin (c\sqrt{K}) \sin (d\sqrt{K}) \cos \delta  = K^2\left( {{X_2} \times {F_2}} \right) \cdot \left( {{X_2} \times {F_1}} \right).
	\end{aligned}
\label{eq:ssc}
\end{eqnarray}
Substituting Eqs.~(\ref{eq:fandx}) and (\ref{eq:sss}) into the numerator and denominator of the left side of Eq.~(\ref{eq:supp-sp-id-2}), we find that
\begin{equation}
\begin{aligned}
	& {\sin (c\sqrt{K}) \sin (d\sqrt{K}) \sin \delta+\sin (a\sqrt{K}) \sin (b\sqrt{K}) \sin \beta} = (y_1-y_2)\sqrt{K}\sin(2f\sqrt{K}),\\
	& {B(\gamma,K)\sin (a\sqrt{K}) \sin (d\sqrt{K}) \sin \alpha+ A(\alpha,K) \sin (b\sqrt{K}) \sin (c\sqrt{K}) \sin \gamma} = (y_1-y_2)\sqrt{K}\sin(2f\sqrt{K}).
\end{aligned}
\end{equation}
\begin{figure}[!t]
\centering
\includegraphics[width=\columnwidth]{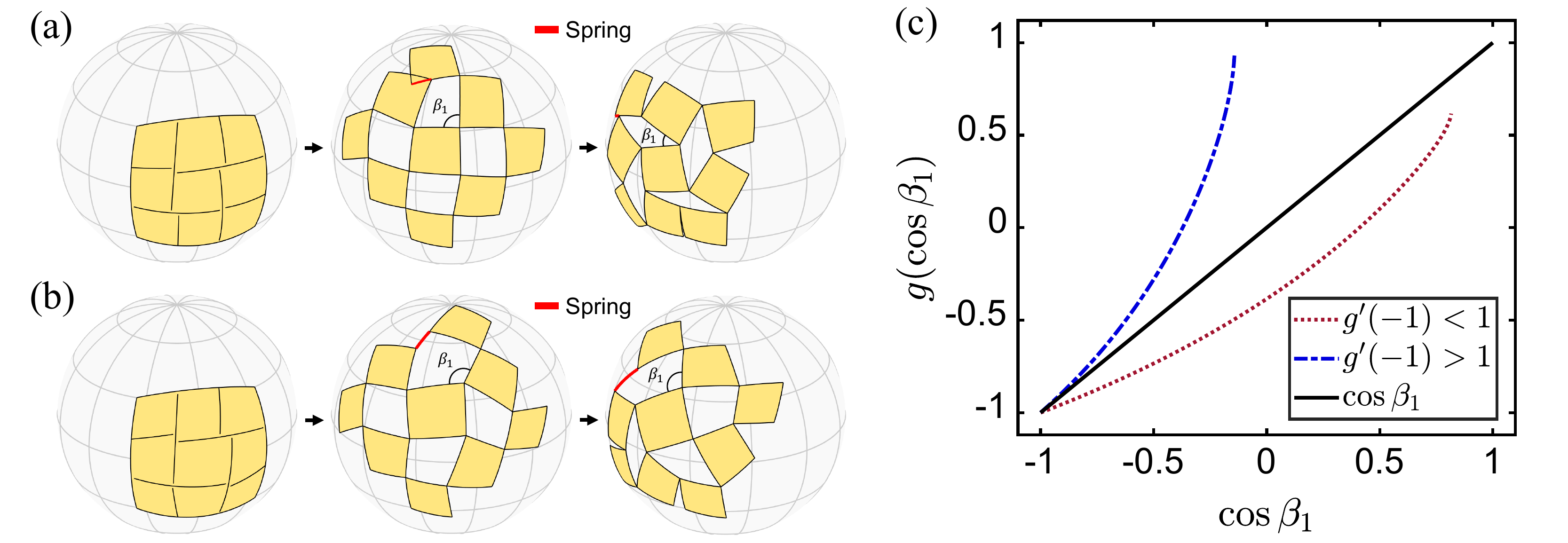}
\caption{The $3\times3$ SQK tessellations with only one compatible state. (a) $g'(-1)<1$; $\cos\beta_1=-1$ (left),~0 (middle),~0.816 (right). (b) $g'(-1)>1$; $\cos\beta_1=-1$ (left),~$-0.5$ (middle),~$-0.143$ (right). (c) Curves of the loop function $g(\cos\beta_1)$.}
\label{fig:SuppSPComp}
\end{figure}
Therefore, Eq.~(\ref{eq:supp-sp-id-2}) is verified.
Now we prove $\bar g''_i > 0$. Substituting Eqs.~(\ref{eq:fandx})--(\ref{eq:ssc}) into Eq.~(\ref{eq:supp-sp-ddg}), we can obtain
\begin{equation}
	\bar g'' = \bar k_g \left(\frac{\sin^2(f\sqrt{K})-\sin^2(l\sqrt{K})}{\sin\theta_1\sin\theta_2}\right)\frac{t^f_1\cos^2(l\sqrt{K})+t^f_2\sin^2(l\sqrt{K})}{t^f_3\cos^2(l\sqrt{K})+t^f_4\sin^2(l\sqrt{K})}.
	\label{eq:supp-sp-ddgp}
\end{equation}
The coefficient functions are
\begin{eqnarray}
	\begin{aligned}
	t^f_1 =& 4\sin^2(f\sqrt{K})\tan^2(f\sqrt{K})(k^f_2\sin\theta_1-k^f_1\sin\theta_2)^2\sin^2\theta_1\sin^2\theta_2-\sin^2(f\sqrt{K})(\sin^2\theta_1-\sin^2\theta_2)^2,\\
	t^f_2 =& \cos ^2(f\sqrt{K}) \left[2 \sin ^2\theta_1-\sin ^2(\theta_1-\theta_2)+2 \sin ^2\theta_2\right] \sin ^2(\theta_1-\theta_2),\\
	t^f_3 =& \sin ^2(f\sqrt{K}) (k^f_2 \sin \theta_1-k^f_1 \sin \theta_2)^2,\\
	t^f_4 =& -\cos ^2(f\sqrt{K}) \sin ^2(\theta_1-\theta_2),
	\end{aligned}
	\label{eq:supp-sp-tfi}
\end{eqnarray}
where $k^f_{1,2} = \sqrt{\cos^2\theta_{1,2}+{\sin^2\theta_{1,2}}/{\cos^2(f\sqrt{K})}}$.
Considering $0<f\sqrt{K}<l\sqrt{K}<\pi/2$, $\theta_1\in(0,\pi)$, $\theta_2\in(-\pi,0)$, we can verify that 
\begin{eqnarray}
    \frac{\sin^2(f\sqrt{K})-\sin^2(l\sqrt{K})}{\sin\theta_1\sin\theta_2}>0.
\end{eqnarray}
\begin{equation}
    t^f_3\cos^2(l\sqrt{K})+ t^f_4\sin^2(l\sqrt{K})= \frac{\sin (a\sqrt{K}) \sin (b\sqrt{K}) \sin (c\sqrt{K}) \sin (d\sqrt{K}) \cos^2(f\sqrt{K}) \sin\alpha \sin\gamma}{\sin^2(l\sqrt{K})-\sin^2(f\sqrt{K})}>0, 
\end{equation}
\begin{equation}
    t^f_2 = \cos ^2(f\sqrt{K}) \left[ \begin{aligned}
    	&2\sin^2 \theta _1\sin^2 \theta _2 + {\left( {\sin {\theta _1} + \sin {\theta _2}} \right)^2}\\
    	&+2\left( {\cos {\theta _1}\cos {\theta _2} - 1} \right)\sin {\theta _1}\sin {\theta _2}
    \end{aligned} \right] \sin ^2(\theta_1-\theta_2)>0,
\end{equation}
\begin{equation}
    t^f_1\cos^2(f\sqrt{K})+t^f_2\sin^2(f\sqrt{K}) = 4 \sin ^2(f\sqrt{K}) \sin ^2\theta_1 \sin ^2\theta_2 \left[\begin{aligned}
    &\cos ^2(f\sqrt{K}) \sin ^2(\theta_1-\theta_2)\\
    +\sin ^2&(f\sqrt{K}) (k^f_2 \sin \theta_1-k^f_1 \sin \theta_2)^2
    \end{aligned}\right]>0,
\end{equation}
\begin{equation}
    t^f_1\cos^2(l\sqrt{K})+t^f_2\sin^2(l\sqrt{K})>\min\left\{t^f_1\cos^2(f\sqrt{K})+t^f_2\sin^2(f\sqrt{K}), t^f_2\right\}>0.
\end{equation}
As a result, we conclude $\bar g''>0$.
Finally, following Eq.~(\ref{eq:supp-sp-ddgi}), we prove $g''_i>0$. 
In the Main Text, this result is used to prove the convexity of the loop function $g$, and to prove that the compatibility condition has at most two roots on the feasible domain $[-1,c_r]$. Examples of $3\times3$ SQK tessellations with two compatible states are provided in the Main Text. Here we supplement two cases with only one compatible state in Fig.~\ref{fig:SuppSPComp}.

For the $3\times3$ SQK tessellations composed of slits with $a_i=c_i$ and $b_i=d_i$, the opening angles satisfy $\alpha_i=\gamma_i$ and $\beta_i=\delta_i$. 
Thus, we have 
\begin{equation}
    2\alpha_i +2\beta_i =2\pi +2K S_{\alpha_i}, 
\end{equation}
and it follows that
\begin{equation}
	\cos \beta_i =-\sin (K S_{\alpha_i}) \sin \alpha_i -\cos (K S_{\alpha_i}) \cos \alpha_i.
	\label{eq:supp-sp-cosBeta_i}
\end{equation}
In this case, the relationships between opening angles of adjacent slits are denoted by $\cos\beta_{i}=g^e_i(\cos\beta_{i+1})$.
Combining Eqs.~(\ref{eq:supp-tanSalpha2})--(\ref{eq:supp-sp-cosSAlpha}), and (\ref{eq:supp-sp-cosBeta_i}), we can obtain
\begin{eqnarray}
	g^e_i(\cos\beta_{i+1})=\frac{\cos\beta_{i+1} [1 + \cos (a_i\sqrt{K}) \cos (b_i\sqrt{K})] -\sin (a_i\sqrt{K}) \sin (b_i\sqrt{K})}{[1 + \cos (a_i\sqrt{K}) \cos (b_i\sqrt{K})] -\cos\beta_{i+1} \sin (a_i\sqrt{K}) \sin (b_i\sqrt{K})}.
	\label{eq:gei}
\end{eqnarray}
Eq.~(\ref{eq:gei}) can also be written in a compact form: 
\begin{equation}
	\label{eq:sp-gei-para}
	g_i^{e}(x)=\frac{(P_i+Q_i)x+(P_i-Q_i)}{(P_i+Q_i)+(P_i-Q_i)x},
\end{equation}
where $P_i=\cos^2[(a_i+b_i)\sqrt{K}/2]$ and $Q_i=\cos^2[(a_i-b_i)\sqrt{K}/2]$.
Then, the composition of $g_i^{e}$ and $g_j^{e}$ can be calculated by 
\begin{equation}
	\label{eq:sp-ge-para}
	g_i^{e} \circ g_j^{e}(x) = \frac{(P_iP_j+Q_iQ_j)x+(P_iP_j-Q_iQ_j)}{(P_iP_j+Q_iQ_j)+(P_iP_j-Q_iQ_j)x}.
\end{equation}
Finally, we obtain the loop function:
\begin{equation}
	\label{eq:sp-para-loop}
	g^{e}(x) = g_1^{e}\circ g_2^{e}\circ g_3^{e}\circ g_4^{e}(x) = \frac{(P+Q) x + (P-Q)}{(P+Q) + (P-Q) x},
\end{equation}
where $P=\prod_{i=1}^4P_i$ and $Q=\prod_{i=1}^4Q_i$.

\section{Formulations of the compatible configurations}

\begin{figure}[!t]
	\centering
	\includegraphics[width=0.8\columnwidth]{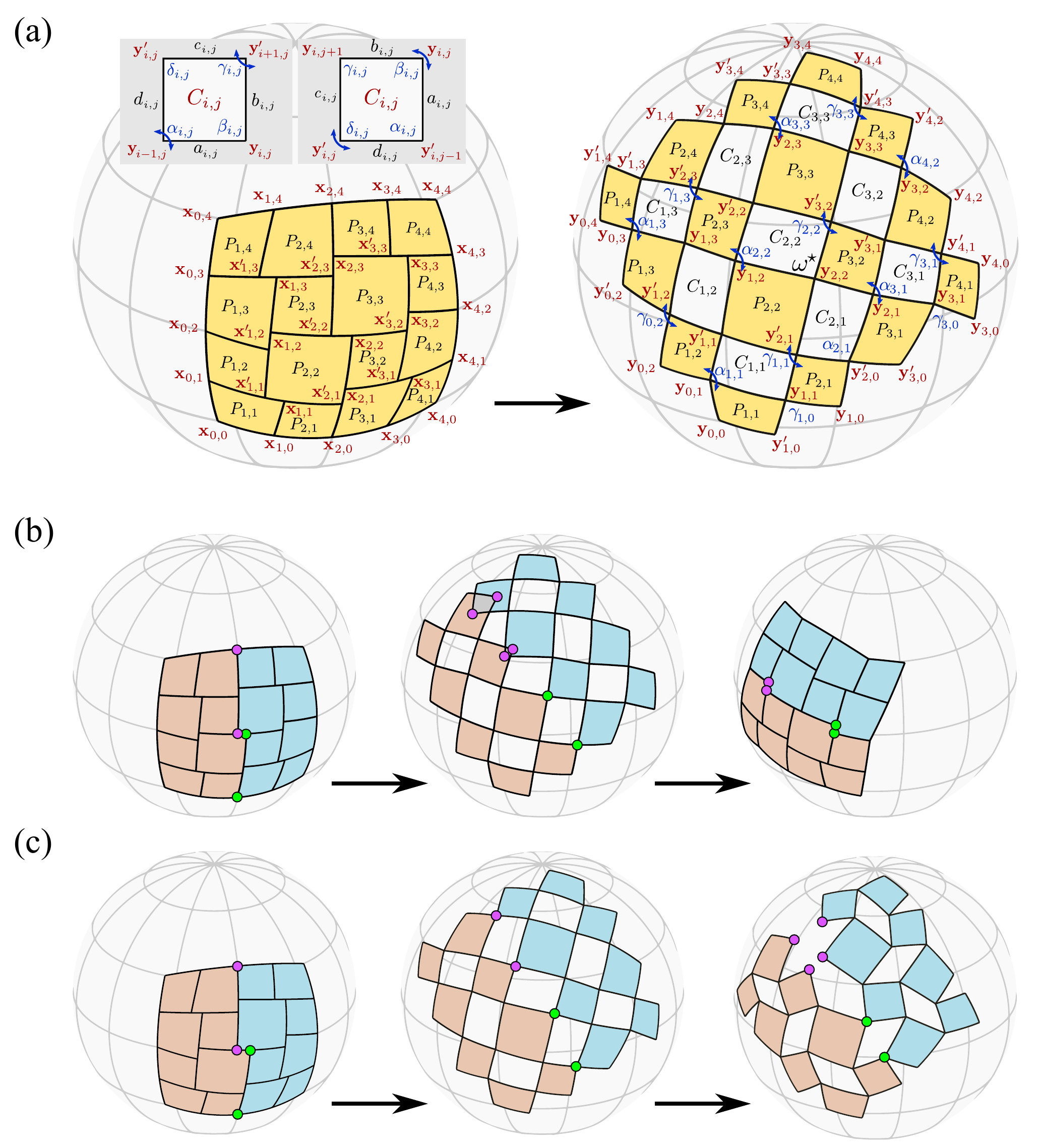}
	\caption{Kinematics of SQK tessellations. (a) Geometric notations of the undeployed configuration and the deployed configuration of a SQK tessellation. The tessellation is compatible at $\cos\omega^\star=-1,0$. (b) The SQK tessellation with $k^b_{i,j}=k^d_{i,j}$ has two compatible states at $\cos\omega^\star=\pm1$. If we disconnect the panels at the purple spots, the tessellation is split into two strips linked at the green spots, and can be rigidly deployed. Each purple spot is divided into two spots (middle), and merge again at $\cos\omega^\star=1$ (right).  (c) The SQK tessellation is optimized to be compatible at $\cos\omega^\star=0$ (middle). The purple spots split apart when $\cos\omega>0$ (right). }
	\label{fig:SuppSPKinematics}
\end{figure}
We formulate the undeployed $M\times N$ SQK patterns by solving a nonlinear equation system under the given boundary conditions and cut ratios.
As illustrated in Fig.~\ref{fig:SuppSPKinematics}(a), the slits of an SQK tessellation are divided by their intersecting neighbors into four segments, so that they form quadrilateral voids when deployed. 
These slits are categorized into horizontal slits [Fig.~\ref{fig:SuppSPKinematics}(a), top left] and vertical slits [Fig.~\ref{fig:SuppSPKinematics}(a), top right], depending on the opening directions indicated by the blue arrows.
The panels are denoted by $P_{i,j}$, and the slits by $C_{i,j}$.
For each slit $C_{i,j}$, we systematically denote the side lengths by $a_{i,j}$, $b_{i,j}$, $c_{i,j}$, $d_{i,j}$, and the opening angles by $\alpha_{i,j}$, $\beta_{i,j}$, $\gamma_{i,j}$, $\delta_{i,j}$.
The two vertices where the slit $C_{i,j}$ is divided are denoted by $\textbf{{x}}_{i,j}$ and $\textbf{{x}}'_{i,j}$.
In addition, the deployed vertices are denoted by $\textbf{{y}}_{i,j}$ and $\textbf{{y}}'_{i,j}$.
To parameterize the shapes of quadrilateral voids, we use $k^b_{i,j}$ and $k^d_{i,j}$ to denote the ratios by which the slit $C_{i,j}$ is divided by the intersecting slits, i.e., $k^b_{i,j}=b_{i,j}/(a_{i,j}+b_{i,j})$ and $k^d_{i,j}=d_{i,j}/(c_{i,j}+d_{i,j})$.
Finally, we define the kinematic parameter $\omega\triangleq\beta_{2,2}$, and use $\omega^\star$ to denote the kinematic parameter at a compatible state.

We establish a Cartesian coordinate system located at the spherical center, so that the positions of vertices can be determined by the spherical coordinates (longitude $\phi$ and latitude $\theta$) as 
\begin{equation}
	\begin{aligned}
    &\textbf{{x}}_{i,j}=(\cos\theta_{i,j}\cos\phi_{i,j}, \cos\theta_{i,j}\sin\phi_{i,j}, \sin\theta_{i,j})/\sqrt{K},\\
    &\textbf{{x}}'_{i,j}=(\cos\theta'_{i,j}\cos\phi'_{i,j}, \cos\theta'_{i,j}\sin\phi'_{i,j}, \sin\theta'_{i,j})/\sqrt{K}.
    \end{aligned}
\end{equation}
We define the function $d_{g}(\textbf{{v}}_1,\textbf{{v}}_2)=\arccos[(\textbf{{v}}_1\cdot\textbf{{v}}_2)/(\|\textbf{{v}}_1\|\|\textbf{{v}}_2\|)]$ to calculate the scaled geodesic distance between two points $\textbf{{v}}_1$ and $\textbf{{v}}_2$ on the sphere.
Since the slits are geodesic lines, the vertices of an $M\times N$ undeployed SQK pattern should satisfy the following nonlinear constraints
\begin{equation}
	\begin{aligned}
		&\tilde f^1_{i,j}(\phi_{k,l},\theta_{k,l},\phi'_{k,l},\theta'_{k,l})\triangleq d_{g}(\textbf{{x}}_{i,j},\textbf{{x}}'_{i+1,j})-k^b_{i,j}d_{g}(\textbf{{x}}_{i-1,j},\textbf{{x}}'_{i+1,j})=0, & \text{mod}(i+j,2)=0,\\
		&\tilde f^2_{i,j}(\phi_{k,l},\theta_{k,l},\phi'_{k,l},\theta'_{k,l})\triangleq d_{g}(\textbf{{x}}'_{i,j},\textbf{{x}}_{i-1,j})-k^d_{i,j}d_{g}(\textbf{{x}}'_{i+1,j},\textbf{{x}}_{i-1,j})=0, & \text{mod}(i+j,2)=0,\\
		&\tilde f^1_{i,j}(\phi_{k,l},\theta_{k,l},\phi'_{k,l},\theta'_{k,l})\triangleq d_{g}(\textbf{{x}}_{i,j},\textbf{{x}}_{i,j+1})-k^b_{i,j}d_{g}(\textbf{{x}}'_{i,j-1},\textbf{{x}}_{i,j+1})=0, & \text{mod}(i+j,2)=1,\\
		&\tilde f^2_{i,j}(\phi_{k,l},\theta_{k,l},\phi'_{k,l},\theta'_{k,l})\triangleq d_{g}(\textbf{{x}}'_{i,j},\textbf{{x}}'_{i,j-1})-k^d_{i,j}d_{g}(\textbf{{x}}_{i,j+1},\textbf{{x}}'_{i,j-1})=0, & \text{mod}(i+j,2)=1,
		\label{eq:supp-sp-EqSys-1}
	\end{aligned}
\end{equation}
\begin{equation}
	\begin{aligned}
		&\hat f^1_{i,j}(\phi_{k,l},\theta_{k,l},\phi'_{k,l},\theta'_{k,l})\triangleq (\textbf{{x}}'_{i+1,j}\times\textbf{{x}}_{i,j})\cdot \textbf{{x}}_{i-1,j} =0, & \text{mod}(i+j,2)=0,\\
		&\hat f^2_{i,j}(\phi_{k,l},\theta_{k,l},\phi'_{k,l},\theta'_{k,l})\triangleq (\textbf{{x}}_{i-1,j}\times\textbf{{x}}'_{i,j})\cdot \textbf{{x}}'_{i+1,j} =0, & \text{mod}(i+j,2)=0,\\
		&\hat f^1_{i,j}(\phi_{k,l},\theta_{k,l},\phi'_{k,l},\theta'_{k,l})\triangleq (\textbf{{x}}_{i,j+1}\times\textbf{{x}}_{i,j})\cdot \textbf{{x}}'_{i,j-1} =0, & \text{mod}(i+j,2)=1,\\
		&\hat f^2_{i,j}(\phi_{k,l},\theta_{k,l},\phi'_{k,l},\theta'_{k,l})\triangleq (\textbf{{x}}'_{i,j-1}\times\textbf{{x}}'_{i,j})\cdot \textbf{{x}}_{i,j+1} =0, & \text{mod}(i+j,2)=1,
		\label{eq:supp-sp-EqSys-2}
	\end{aligned}
\end{equation}
for $i=1,...,M-1$ and $j=1,...,N-1$. 
Eq.~(\ref{eq:supp-sp-EqSys-1}) describes the constraints of side-length ratios, and Eq.~(\ref{eq:supp-sp-EqSys-2}) restricts that vertices of the same slit are on the same great circle.
Fixing the boundary vertices of the tessellation, i.e., $\textbf{{x}}^L_{j}$, $\textbf{{x}}^R_{j}$, $\textbf{{x}}^B_{i}$, and $\textbf{{x}}^U_{i}$ for the left, right, bottom, and upper boundaries, respectively, we can write the boundary conditions for Eqs.~(\ref{eq:supp-sp-EqSys-1}) and (\ref{eq:supp-sp-EqSys-2}) as
\begin{equation}
	\begin{aligned}
		&\textbf{{x}}_{0,j}=\textbf{{x}}'_{0,j}=\textbf{{x}}^L_{j}, && \text{mod}(i+j,2)=0,\\
		&\textbf{{x}}_{M,j}=\textbf{{x}}'_{M,j}=\textbf{{x}}^R_{j}, && \text{mod}(i+j,2)=0,\\
		&\textbf{{x}}_{i,0}=\textbf{{x}}'_{i,0}=\textbf{{x}}^B_{i}, && \text{mod}(i+j,2)=1,\\
		&\textbf{{x}}_{i,N}=\textbf{{x}}'_{i,N}=\textbf{{x}}^U_{i}, && \text{mod}(i+j,2)=1,
		\label{eq:supp-sp-EqSysBC}
	\end{aligned}
\end{equation}
for $i=0,1,...,M$ and $j=0,1,...,N$.
Given the cut ratios $k^b_{i,j}$ and $k^d_{i,j}$, there are two equations and two unknowns (the $\phi$ and $\theta$ coordinates) for each interior vertex $\textbf{{x}}_{i,j}$ or $\textbf{{x}}'_{i,j}$ according to Eqs.~(\ref{eq:supp-sp-EqSys-1}) and (\ref{eq:supp-sp-EqSys-2}), so that the nonlinear equation system is closed.
We use the function {\bf fmincon} in Matlab R2020b to solve Eqs.~(\ref{eq:supp-sp-EqSys-1})--(\ref{eq:supp-sp-EqSysBC}).
As a result, we can obtain the positions of the interior vertices $\textbf{{x}}_{i,j}$ and $\textbf{{x}}'_{i,j}$, and consequently the SQK pattern.

In the Main Text, we have proved that the SQK tessellations with $k^b_{i,j}=k^d_{i,j}$ have two compatible states at $\cos\omega^\star=\pm1$.
In what follows, we will shift the latter compatible state from $\cos\omega^\star=1$ into $\cos\omega^\star\in(-1,1)$, and construct corresponding SQK tessellations. 
Firstly, we relax the constraints by disconnecting links between some panels, such that the tailored tessellations become rigidly deployable.
For example, Fig.~\ref{fig:SuppSPKinematics}(b) illustrates the disconnected links (purple spots) of a $4\times4$ square SQK tessellation with $k^b_{i,j}=k^d_{i,j}=0.45$. 
In general, to relax an $M\times N$ tessellation, we disconnect adjacent panels at the vertices $\textbf{{x}}_{2p,2q}$ ($p\geq1,q\geq2$) and $\textbf{{x}}'_{2p,2q}$ ($p\geq1,q\geq1$).
We use $\mathcal{R}_{i,j}$ to denote the rigid transformations of the panels $P_{i,j}$ relative to the panel $P_{1,1}$.
The transformations $\mathcal{R}_{i,j}$ can be calculated by iteratively superimpose the rotation of a panel over its neighbor on the bottom ($j>1$), or on the left ($j=1$).
This procedure can be formulated as follows:
\begin{equation}
	\begin{aligned}
		&\mathcal{R}_{1,1}=\mathcal{I}, && i=1,~j=1,\\
		&\mathcal{R}_{i,1}=\textbf{{R}}(\gamma_{i-1,0},\mathcal{R}_{i-1,1}\textbf{{x}}_{i-1,1})\mathcal{R}_{i-1,1}, && \text{mod}(i+1,2)=1,~j=1,\\
		&\mathcal{R}_{i,1}=\textbf{{R}}(-\alpha_{i-1,1},\mathcal{R}_{i-1,1}\textbf{{x}}'_{i-1,0})\mathcal{R}_{i-1,1}, && \text{mod}(i+1,2)=0,~j=1,\\
		&\mathcal{R}_{i,j}=\textbf{{R}}(\alpha_{i,j-1},\mathcal{R}_{i,j-1}\textbf{{x}}_{i-1,j-1})\mathcal{R}_{i,j-1}, && \text{mod}(i+j,2)=1,~j>1,\\
		&\mathcal{R}_{i,j}=\textbf{{R}}(-\gamma_{i-1,j-1},\mathcal{R}_{i,j-1}\textbf{{x}}'_{i,j-1})\mathcal{R}_{i,j-1}, && \text{mod}(i+j,2)=0,~j>1,
		\label{eq:supp-sp-trans}
	\end{aligned}
\end{equation}
where $\mathcal{I}$ is the identity transformation, and $\textbf{{R}}(\varphi,\textbf{{x}}_0)$ is the rotation transformation around $\textbf{{x}}_0$ on the sphere, defined by
\begin{equation}
	\textbf{{R}}(\varphi,\textbf{{x}}_0)\textbf{{x}}:=\frac{\textbf{{x}}_0(\textbf{x}_0\cdot\textbf{{x}})}{\|\textbf{{x}}_0\|^2} +\frac{(\textbf{{x}}_0\times\textbf{{x}})\times\textbf{{x}}_0}{\|\textbf{{x}}_0\|^2}\cos\varphi +\frac{\textbf{{x}}_0\times\textbf{{x}}}{\|\textbf{{x}}_0\|}\sin\varphi.
\end{equation}
For a fixed kinematic parameter $\omega\triangleq\beta_{2,2}$, the opening angles $\alpha_{i,j}, \gamma_{i,j}$ in Eq.~(\ref{eq:supp-sp-trans}) should be solved according to Eqs.~(\ref{eq:supp-sp-cos-1}), (\ref{eq:supp-sp-cos-2}), (\ref{eq:cb}), (\ref{sp-cos-1}), and the following conserved relations:
\begin{equation}
	\alpha_{i,j}=\pi-\beta_{i-1,j},~
	\gamma_{i,j}=\pi-\delta_{i+1,j},~
	\beta_{i,j}=\pi-\gamma_{i,j-1},~
	\delta_{i,j}=\pi-\alpha_{i,j+1},~\text{mod}(i+j,2)=0.
	\label{eq:supp-sp-complementary}
\end{equation}
Then, the displacement field of the tailored SQK tessellation can be written as
\begin{equation}
	\textbf{{y}}=\mathcal{R}_{i,j}\textbf{{x}},~\textbf{{x}}\in P_{i,j}.
	\label{eq:supp-sp-deformation}
\end{equation}
Now we can write the conditions for a SQK tessellation to be compatible at $\cos\omega^\star\in(-1,1)$:
\begin{equation}
	\begin{aligned}
		&\tilde{h}^1_{p,q}(\phi_{k,l},\theta_{k,l},\phi'_{k,l},\theta'_{k,l},\omega^\star)\triangleq
		K\|\mathcal{R}_{2p,2q}\textbf{{x}}_{2p,2q}-\mathcal{R}_{2p+1,2q}\textbf{{x}}_{2p,2q}\|^2=0, && p\geq1,q\geq2,\\
		&\tilde{h}^2_{p,q}(\phi_{k,l},\theta_{k,l},\phi'_{k,l},\theta'_{k,l},\omega^\star)\triangleq
		K\|\mathcal{R}_{2p,2q+1}\textbf{{x}}'_{2p,2q}-\mathcal{R}_{2p+1,2q+1}\textbf{{x}}'_{2p,2q}\|^2=0, && p\geq1,q\geq1.
		\label{eq:supp-sp-comp}
	\end{aligned}
\end{equation}
Eq.~(\ref{eq:supp-sp-comp}) means that the disconnected panels are linked again at $\cos\omega^\star$, as illustrated in Fig.~\ref{fig:SuppSPKinematics}(c).
To determine the SQK patterns compatible at $\cos\omega^\star\in(-1,1)$, we start from an initial pattern with $k^b_{i,j}=k^d_{i,j}=\bar{k}_{i,j}$ (i.e., compatible at $\cos\omega^\star=1$), then optimize to achieve Eq.~(\ref{eq:supp-sp-comp}). To obtain an optimized pattern close to the input one, we solve the following optimization problem
\begin{equation}
	\min_{\phi_{k,l},\,\theta_{k,l},\,\phi'_{k,l},\,\theta'_{k,l}} 
	\begin{cases}
		(k^b_{i,j}-\bar{k}_{i,j})^2,\\
		(k^d_{i,j}-\bar{k}_{i,j})^2,
	\end{cases}
	\text{subject to}
	\begin{cases}
		\tilde{h}^m_{p,q}(\phi_{k,l},\theta_{k,l},\phi'_{k,l},\theta'_{k,l},\omega^\star)=0,\\
		\hat{f}^m_{i,j}(\phi_{k,l},\theta_{k,l},\phi'_{k,l},\theta'_{k,l})=0,
	\end{cases}
	\label{eq:supp-sp-opt-loop}
\end{equation}
for $m=1,2$.
We add the constraints $\hat{f}^m_{i,j}=0$ [Eq.~(\ref{eq:supp-sp-EqSys-2})] to preserve that vertices of the same slit stay on the same great circle during optimization.
The boundary vertices are fixed according to Eq.~(\ref{eq:supp-sp-EqSysBC}) in the optimization, so that the corresponding spherical coordinates are not included in the optimization variables.
We use the function {\bf fgoalattain} in Matlab R2020b to solve Eq.~(\ref{eq:supp-sp-opt-loop}).

\section{Planar tessellations}

If the side lengths of SQK tessellations tend to zero on the unit sphere, the compatibility condition is degenerated to characterize the planar quadrilateral kirigami (PQK) tessellations. Specifically, by applying $K \to 0$, we can obtain the planar forms of Eqs.~(\ref{eq:supp-sp-dg}) and (\ref{eq:supp-sp-ddgp}):
\begin{equation}
	\tilde g'(\cos\alpha;a,b,c,d) = -\frac{ d \sin \beta \sin \delta}{ b \sin \alpha \sin \gamma},
	\label{eq:supp-pl-dg}
\end{equation}
\begin{equation}
	\tilde g'' = \tilde k_g \left(\frac{f^2-l^2}{\sin\theta_1\sin\theta_2}\right)\frac{\tilde t^f_1+\tilde t^f_2 l^2}{\tilde t^f_3+\tilde t^f_4 l^2},
	\label{eq:supp-pl-ddgp}
\end{equation}
where the coefficient $\tilde k_g$ is
\begin{equation}
    \tilde k_g=\frac{d\sin\beta\sin\delta}{b^2c\sin^2\alpha\sin^2\gamma}>0.
\end{equation}
The coefficient functions are given by
\begin{eqnarray}
	\begin{aligned}
		\tilde t^f_1 =& -f^2(\sin^2\theta_1-\sin^2\theta_2)^2,\\
		\tilde t^f_2 =& \left[2 \sin ^2\theta_1-\sin ^2(\theta_1-\theta_2)+2 \sin ^2\theta_2\right] \sin ^2(\theta_1-\theta_2),\\
		\tilde t^f_3 =& f ^2 (\sin \theta_1-\sin \theta_2)^2,\\
		\tilde t^f_4 =& -\sin ^2(\theta_1-\theta_2).
	\end{aligned}
	\label{eq:supp-pl-tfi}
\end{eqnarray}
Now suppose that the quadrilateral is not a parallelogram (i.e., $\theta_1-\theta_2 \neq \pi$).
Since $l>f$ and $\tilde t^f_2>0$, we have 
\begin{equation}
    \tilde t^f_1+\tilde t^f_2 l^2>\tilde t^f_1+\tilde t^f_2 f^2=4f^2\sin ^2\theta_1\sin ^2(\theta_1-\theta_2)\sin ^2\theta_2>0.
\end{equation}
\begin{equation}
    \tilde t_3^f + \tilde t_4^f  l^2=\frac{a b c d \sin\alpha \sin\gamma}{l^2-f^2}>0. 
\end{equation}
Consequently, we prove $\tilde g''>0$ for a non-parallelogram quadrilateral void of PQK tessellations. 
Analogous to the spherical case, this result leads to the strict convexity of the loop function, so that PQK tessellations with non-parallelogram slits also have either one or two compatible configurations.

\section{Energetics}

\subsection{Single-spring model}

\begin{figure}[!t]
	\centering
	\includegraphics[width=0.8\columnwidth]{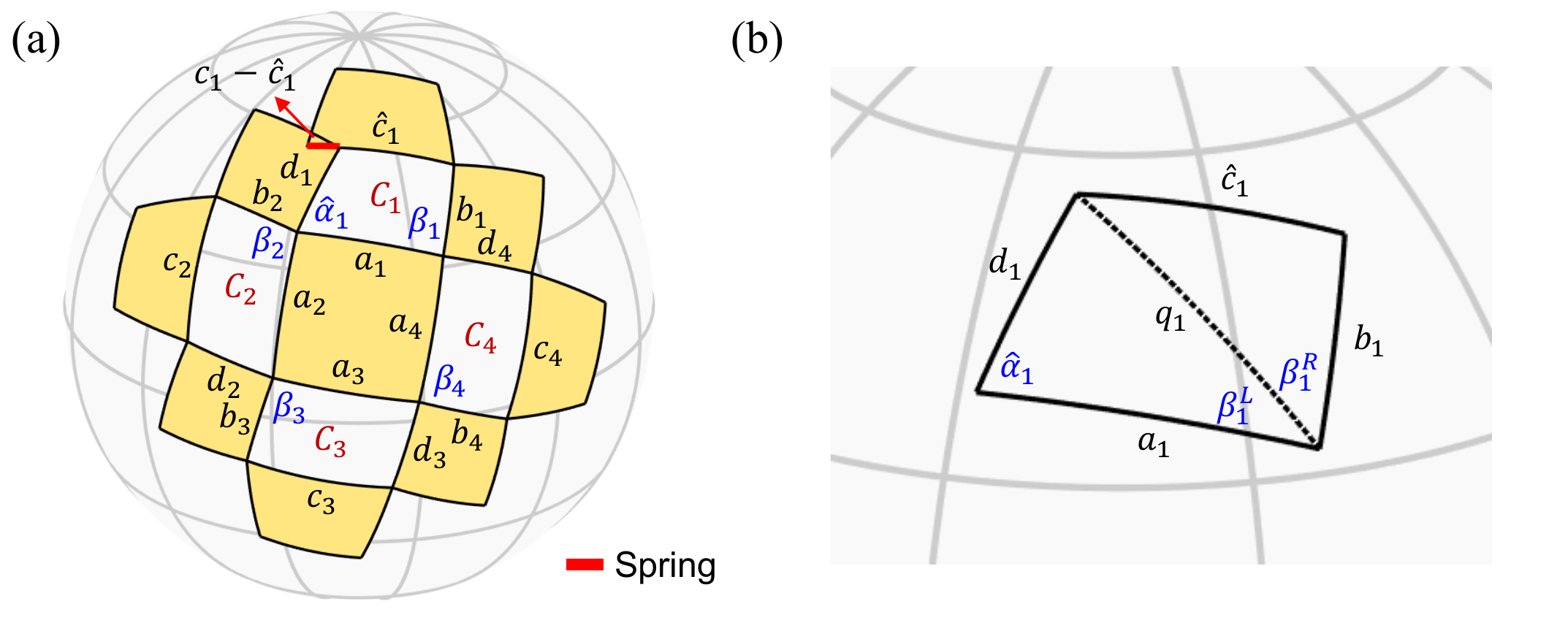}
	\caption{The single-spring model. (a) The cut-off vertices are linked by a zero-length spring. In this way, at any incompatible state of the $3\times3$ SQK tessellation, the spring always holds the shortest length to achieve minimum potential energy. (b) The slit $C_1$ extracted from the tessellation in (a). The diagonal $q_1$ divides the opening angle $\beta_1$ into $\beta_1^L$ and $\beta_1^R$.}
	\label{fig:SuppSpring}
\end{figure}

Here we present a single-spring model to simulate the energy landscapes upon the deployment of a $3\times3$ SQK tessellation.
in this model, the panels are rigid and connected by revolute joints, except that one specific joint is replaced by a spring.
As shown in Fig.~\ref{fig:SuppSpring}(a), we disconnect the vertex above slit $C_1$, and link the split vertices by a zero-length spring with stiffness $k_S$.
At an incompatible state, the spring will keep shortest elongation to minimize the elastic energy.
Thus, the spring and the bottom edge of the flexible panel are always collinear, i.e., on the same great circle.
Then, the elongation of the spring is $\Delta_S=|c_1-\bar c_1|$, where $\hat c_1$ is the top side length of the slit $C_1$, and $c_1$ the bottom edge length of the flexible panel.

In the single-spring model, the elastic energy of a $3\times3$ SQK tessellation is given by
\begin{equation}
	E_S=(k_S/2)(c_1-\hat c_1)^2.
\end{equation}
As shown in Fig.~\ref{fig:SuppSpring}(b), the diagonal $q_1$ of the slit $C_1$ divides $\beta_1$ into two angles $\beta_1^L$ and $\beta_1^R$.
Following spherical trigonometry, we have
\begin{equation}
	\cos(q_1\sqrt{K}) = \cos(a_1\sqrt{K}) \cos(d_1\sqrt{K}) + \sin(a_1\sqrt{K}) \sin(d_1\sqrt{K}) \cos\hat\alpha_1,
	\label{eq:cosq1sk}
\end{equation}
\begin{equation}
	\sin\beta_1^L = \sin\hat\alpha_1 \frac{\sin(d_1\sqrt{K})}{\sin(q_1\sqrt{K})},
\end{equation}
\begin{equation}
	\cos\beta_1^L = \frac{\cos(d_1\sqrt{K}) - \cos(a_1\sqrt{K})\cos(q_1\sqrt{K})}{\sin(a_1\sqrt{K})\sin(q_1\sqrt{K})},
	\label{eq:cosbeta1L}
\end{equation}
\begin{equation}
	\cos(\hat c_1\sqrt{K}) = \cos(q_1\sqrt{K}) \cos(b_1\sqrt{K}) + \sin(q_1\sqrt{K}) \sin(b_1\sqrt{K}) \cos\beta_1^R.
	\label{eq:cosc1hat}
\end{equation}
Since $\cos\beta_1^R=\cos\beta_1\cos\beta_1^L+\sin\beta_1\sin\beta_1^L$, we substitute Eqs.~(\ref{eq:cosq1sk})--(\ref{eq:cosbeta1L}) into Eq.~(\ref{eq:cosc1hat}), and obtain
\begin{equation}
	\begin{aligned}	
	&\hat c_1 = \frac{1}{\sqrt{K}}\arccos\left\{
	\begin{aligned}	
	& \sin (b_1 \sqrt{K}) \sin (d_1 \sqrt{K}) \left[\sin \hat\alpha_1 \sin \beta_1-\cos \hat\alpha_1 \cos \beta_1 \cos (a_1 \sqrt{K})\right]\\
	& +\sin (a_1 \sqrt{K}) \sin (b_1 \sqrt{K}) \cos(d_1 \sqrt{K}) \cos \beta_1\\
	& +\cos (b_1 \sqrt{K}) \left[\cos \hat\alpha_1 \sin (a_1 \sqrt{K}) \sin (d_1 \sqrt{K})+\cos (a_1 \sqrt{K}) \cos (d_1 \sqrt{K})\right]
    \end{aligned}	
	\right\},\\
	&\hat\alpha_1 = \arccos[g_2\circ g_3 \circ g_4(\cos\beta_1)],
    \end{aligned}
\label{eq:cosc1hatsk}
\end{equation}
where $g_2$, $g_3$, $g_4$ are given by Eq.~(\ref{eq:supp-sp-gi}). 
Eq.~(\ref{eq:cosc1hatsk}) gives the explicit expression of $\hat c_1$ with respect to $\cos\beta_1$.
Therefore, the elastic energy $E_S$ is explicitly determined by the kinematic parameter $\cos\beta_1$.

Now we derive the approximate formulas of the energy $E_S$ when $K$ is small for $3\times3$ SQK tessellations with equal opposite side lengths of slits, i.e., $c_i=a_i$ and $d_i=b_i$.
To be clear, a small $K$ satisfies $L\sqrt{K}\ll1$ for $L=\max\{a_1,b_1,...,a_4,b_4\}$.
In this case, we calculate the Taylor series of ${\hat c_1}/{b_1}$ following Eq.~(\ref{eq:cosc1hatsk}): 
\begin{equation}
	\frac{\hat c_1}{b_1} = \frac{c_1}{b_1} - \frac{1}{2}\sum_{i=1}^4(a_ib_i)K\sin^2\beta_1 + O[L^4K^2].
	\label{eq:hatc1b1}
\end{equation}
Thus, the scaled energy $E_S/(k_Sb_1^2)$ can be approximated by
\begin{equation}
	\frac{E_S}{k_Sb_1^2} = \frac{1}{8}\left[\sum_{i=1}^4(a_ib_i)\right]^2K^2\sin^4\beta_1 + O[L^6K^3].
	\label{eq:es}
\end{equation}
Additionally, we can calculate the Taylor series of the loop function $g^{e}(\cos\beta_1)$ based on Eq.~(\ref{eq:sp-para-loop}):
\begin{equation}
	g^{e}(\cos\beta_1) = \cos\beta_1 - \frac{1}{2}\sum_{i=1}^4(a_ib_i)K\sin^2\beta_1 + O[L^4K^2].
	\label{eq:gcosbeta1}
\end{equation}
Comparing Eqs.~(\ref{eq:hatc1b1}) and (\ref{eq:gcosbeta1}), we obtain
\begin{equation}
	\frac{\Delta_S}{b_1} = \frac{|\hat c_1-c_1|}{b_1} = |g^{e}(\cos\beta_1) - \cos\beta_1| + O[L^4K^2].
\end{equation}
Finally, we obtain the scaled energy expressed by the loop function:
\begin{equation}
	\frac{E_S}{k_Sb_1^2} = \frac{1}{2} [g^{e}(\cos\beta_1) - \cos\beta_1]^2 + O[L^6K^3].
\end{equation}

\begin{figure}[!t]
\centering
\includegraphics[width=\textwidth]{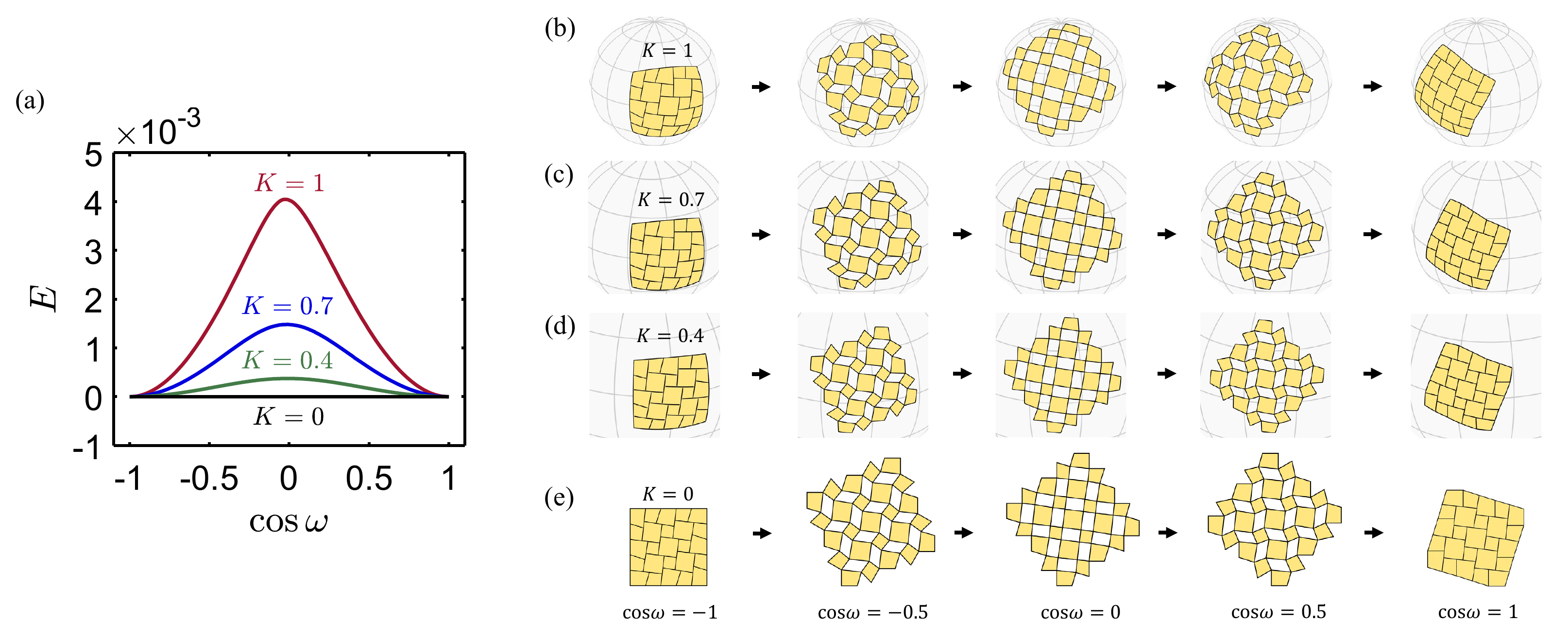}
\caption{(a) The evolution of elastic energy $E$ of $5\times5$ SQK and PQK shells with equal opposite side lengths of slits in respect of the kinematic parameter $\cos\omega$. The square shells have the same side length $s=\pi/3$ but decreasing Gaussian curvatures $K=1$, $0.7$, $0.4$, and $0$. The aspect ratios of the slits are fixed as 0.4. (b)--(d) The deployment path of the SQK shells. The configurations for different $\cos\omega$ are determined by minimizing the elastic energy of the multispring model. (e) The deployment path of the PQK shell.}
\label{fig:SuppSPKappa}
\end{figure}

\subsection{Multispring model}

For a general $M\times N$ SQK tessellation, we assume the hinge-connected panels are made of elastic materials.
To simulate the deformation of the panels, we use a multispring model, in which the vertices are linked by springs along the edges and diagonals of panels.
Then, the elastic energy of the deployed tessellations can be written as
\begin{equation}
	\begin{aligned}
	&E\left(\textbf{{Y}}\right) = \sum_{n}\frac{k_n}{2}\left[l_n\left(\textbf{{Y}}\right)-l^0_n\right]^2,\\
	&\textbf{{Y}}=\left\{\textbf{{y}}_{i,j},\textbf{{y}}'_{i,j}|i=0,1,...,M, j=0,1,...,N\right\},
	\end{aligned}
	\label{eq:supp-sp-energy}
\end{equation}
where $\textbf{{Y}}$ is the array consisting of all the vertex positions of panels, $l_n$ the spring length indexed by $n$, $l^0_n$ the rest spring length for the undeployed tessellation, and $k_n$ the spring stiffness.
For simplicity, we set $k_n=1/l^0_n$ in our demonstration.
We define the function
\begin{equation}
	\eta(\textbf{{v}}_1,\textbf{{v}}_2,\textbf{{v}}_3)= \frac{\textbf{{v}}_1\times\textbf{{v}}_2}{\|\textbf{{v}}_1\times\textbf{{v}}_2\|}\cdot\frac{\textbf{{v}}_3\times\textbf{{v}}_2}{\|\textbf{{v}}_3\times\textbf{{v}}_2\|}
\end{equation}
so that the kinematic parameter $\cos\omega$ can be expressed as $\cos\omega\triangleq\cos\beta_{2,2}=\eta(\textbf{{y}}_{1,2},\textbf{{y}}_{2,2},\textbf{{y}}'_{3,2})$. We use the spherical coordinates (longitude $\tilde\phi$ and latitude $\tilde\theta$) to represent the vertices
\begin{eqnarray}
\textbf{{y}}_{i,j} &=& (\cos\tilde\theta_{i,j}\cos\tilde\phi_{i,j}, \cos\tilde\theta_{i,j}\sin\tilde\phi_{i,j}, \sin\tilde\theta_{i,j})/\sqrt{K},\\
\textbf{{y}}'_{i,j} &=& (\cos\tilde\theta'_{i,j}\cos\tilde\phi'_{i,j}, \cos\tilde\theta'_{i,j}\sin\tilde\phi'_{i,j}, \sin\tilde\theta'_{i,j})/\sqrt{K}.
 \end{eqnarray}
To determine the vertex coordinates of deformed SQK tessellations, we start from the undeployed state at $\cos\omega=-1$, increase $\cos\omega$ incrementally, and minimize the elastic energy Eq.~(\ref{eq:supp-sp-energy}) at each step by optimizing the positions of the vertices $\textbf{{y}}_{i,j}$ and $\textbf{{y}}'_{i,j}$ on the sphere.  We use the function {\bf fmincon} in Matlab R2020b to solve the following optimization problem
\begin{equation}
	\min_{\tilde\phi_{k,l},\tilde\theta_{k,l},\tilde\phi'_{k,l},\tilde\theta'_{k,l}} E\left(\textbf{{y}}_{i,j},\textbf{{y}}'_{i,j}\right) \text{subject to}
	\begin{cases}
		[\eta(\textbf{{y}}_{1,2},\textbf{{y}}_{2,2},\textbf{{y}}'_{3,2})-\cos\omega]^2=0,\\
		{{{({{\tilde \phi }_{1,1}} - {{\bar \phi }_{1,1}})}^2} = 0,{\rm{ }}{{({{\tilde \theta }_{1,1}} - {{\bar \theta }_{1,1}})}^2} = 0,{\rm{ }}{{({{\tilde \phi }_{1,2}} - {{\bar \phi }_{1,2}})}^2} = 0}.
	\end{cases}
	\label{eq:supp-pl-opt-energy}
\end{equation}
In this optimization, we use $(\bar\phi_{0,0},\bar\theta_{0,0},\bar\phi_{0,1})$ to express the orientation of the tessellation on the sphere.

\begin{figure}[!t]
	\centering
	\includegraphics[width=0.8\textwidth]{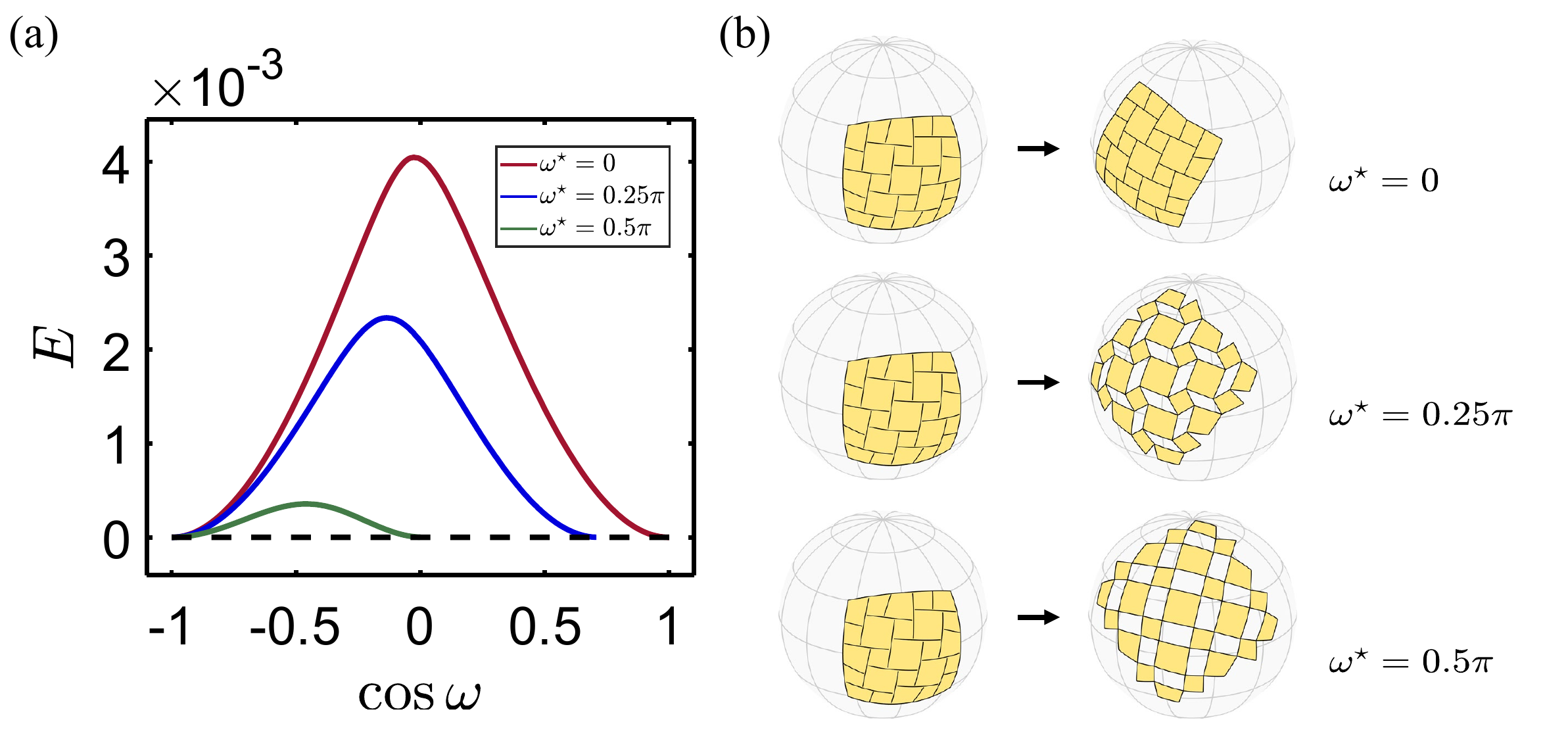}
	\caption{(a) The evolution of elastic energy $E$ of $5\times5$ SQK tessellations as a function of the kinematic parameter $\cos\omega$. (b) The SQK tessellations have the same side length $s=\pi/3$ and Gaussian curvature $K=1$. They are designed to be compatible at $\cos\omega=\cos\omega^\star=0,~1/\sqrt{2}$, and $1$, respectively.}
	\label{fig:SuppEnergy}
\end{figure}

In Fig.~\ref{fig:SuppSPKappa}, we illustrate the energy evolution upon deployment of square SQK and PQK tessellations of the same side length but different Gaussian curvatures $K=1,~0.7,~0.4$, and $0$.
The configurations of these tessellations on the deployment path are also provided.
In Fig.~\ref{fig:SuppEnergy}, we illustrate the energy evolution upon deployment of square SQK tessellations designed to be compatible at $\cos\omega=\cos\omega^\star=0,~1/\sqrt{2}$, and $1$, respectively. 
One can observe that each energy curve has two zero points. 
Moreover, the energy barrier is lower if the two compatible states of a tessellation are closer to each other.

\section{Inverse design}

\begin{figure}[!t]
	\centering
	\includegraphics[width=\columnwidth]{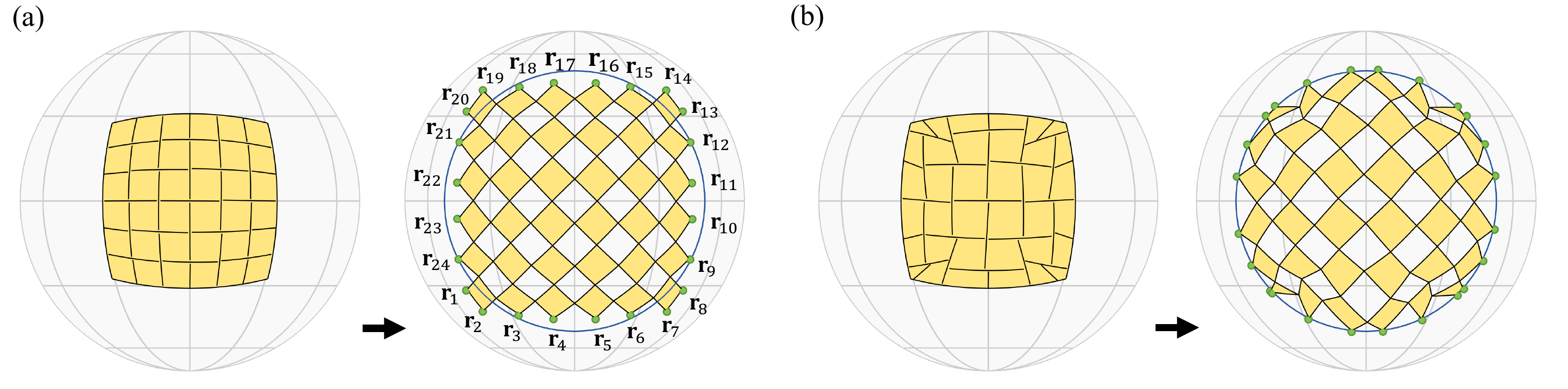}
	\caption{Shape morphing of a square SQK tessellation. (a) The square SQK tessellation with side length $s=0.3\pi$ and Gaussian curvature $K=1$ is deployed ($\omega^\star=0.5\pi$) to align with a small circle at $x=0.65$. (b) The optimized SQK tessellation approximates the circle.}
	\label{fig:SuppSPShape}
\end{figure}

In this section, we aim to optimize a square SQK tessellation on a unit sphere to achieve a domelike shape at a certain deployed state.  
The design problem can be formulated by minimizing the distance between the target curve and outer vertices at the deployed state over the domain of vertex locations on the sphere.  
For example, consider an $N\times N$ square SQK pattern ($N=6$, $k_{i,j}\approx k'_{i,j}\approx 0.5$) with side length $s$.
the corners of the square can be given by
\begin{equation}
	\begin{aligned}
		&\textbf{{x}}^{LB} = ( \cos\theta_0\cos\phi_0, ~-\cos\theta_0\sin\phi_0, ~-\sin\theta_0),\\
		&\textbf{{x}}^{RB} = ( \cos\theta_0\cos\phi_0, ~\cos\theta_0\sin\phi_0, ~-\sin\theta_0),\\
		&\textbf{{x}}^{RU} = ( \cos\theta_0\cos\phi_0, ~\cos\theta_0\sin\phi_0, ~ \sin\theta_0),\\
		&\textbf{{x}}^{LU} = ( \cos\theta_0\cos\phi_0, ~-\cos\theta_0\sin\phi_0, ~ \sin\theta_0),
		\label{eq:supp-sp-BVC}
	\end{aligned}
\end{equation}
where $\theta_0=s/2$ and $\phi_0=\arcsin(\tan(s/2))$.
The boundary vertices are determined by four series of parameters:
\begin{equation}
	\begin{aligned}
		&\textbf{{x}}^B_{i} = \textbf{{R}}(t_i^Bs,\textbf{{x}}^{LB}\times\textbf{{x}}^{RB})\textbf{{x}}^{LB},\\
		&\textbf{{x}}^U_{i} = \textbf{{R}}(t_i^Us,\textbf{{x}}^{LU}\times\textbf{{x}}^{RU})\textbf{{x}}^{LU},\\
		&\textbf{{x}}^L_{i} = \textbf{{R}}(t_i^Ls,\textbf{{x}}^{LB}\times\textbf{{x}}^{LU})\textbf{{x}}^{LB},\\
		&\textbf{{x}}^R_{i} = \textbf{{R}}(t_i^Rs,\textbf{{x}}^{RB}\times\textbf{{x}}^{RU})\textbf{{x}}^{RB},
		\label{eq:supp-sp-BV}
	\end{aligned}
\end{equation}
where $i=0,1,...,N$, $0=t_0^{ I}<t_1^{ I}<...<t_N^{ I}=1$, ${ I}={ L,R,B,U}$.
The outer vertices at the deployed compatible state are denoted by $\textbf{{r}}_n=(x_n,y_n,z_n)$, for $n=1,2,...,4N$, highlighted by green dots in Fig.~\ref{fig:SuppSPShape}(a). 
The equation of a circle on the unit sphere can be given by $\tilde{g}(\textbf{{r}},\lambda)=x-\lambda$ for $\textbf{{r}}=(x,y,z)$. 
We use the parameter $\lambda\in(-1,1)$ to control the size and location of the target circle. 
To characterize the distance between the target circle and the outer vertices $\textbf{{r}}_n$ for a fixed kinematic parameter $\omega=\omega^\star$, we define the following function:
\begin{equation}
    \tilde{g}_{i,j}(\phi_{k,l},\theta_{k,l},\phi'_{k,l},\theta'_{k,l},\lambda,\textbf{{R}},\omega^\star)=[\tilde{g}(\textbf{{R}}\textbf{{r}}_m,\lambda)]^2, 
\end{equation}
where $\phi_{k,l},\theta_{k,l},\phi'_{k,l},\theta'_{k,l}$ are spherical coordinates of the vertices $\textbf{{x}}_{i,j}$ and $\textbf{{x}}'_{i,j}$ on the undeployed tessellation, and $\textbf{{R}}=\textbf{{R}}(\varphi,\textbf{{x}}_0)$ the rotation transformation controlling the orientation of the tessellation. 
We approximate a circle with $\lambda=\lambda_0$ by tackling the following optimization problem:
\begin{equation}
	\min_{\phi_{k,l},\,\theta_{k,l},\,\phi'_{k,l},\,\theta'_{k,l},~ t_k^I,~\lambda,~\textbf{{R}}} (\lambda-\lambda_0)^2 ~\text{subject to}~
	\begin{cases}
		\tilde{g}_{i,j}(\phi_{k,l},\theta_{k,l},\phi'_{k,l},\theta'_{k,l},\lambda,\textbf{{R}},\omega^\star)=0,\\
		\tilde{h}^m_{p,q}(\phi_{k,l},\theta_{k,l},\phi'_{k,l},\theta'_{k,l},\omega^\star)=0,\\
		0=t_0^{I}< t_1^{I}<...< t_N^{I}=1.
	\end{cases}
	\label{eq:sp-opt-shape}
\end{equation}
The constraints $\hat{f}^m_{i,j}=0$ restrict that the vertices of the same slits locating on the same great circle [Eq.~(\ref{eq:supp-sp-EqSys-2})]. The constraints $\tilde{h}^m_{p,q}=0$ preserve the compatibility [Eq.~(\ref{eq:supp-sp-comp})]. 
We use the function $\textbf{fgoalattain}$ in Matlab R2020b to solve Eq.~(\ref{eq:sp-opt-shape}).
The optimized tessellation approximating a small circle is illustrated in Fig.~\ref{fig:SuppSPShape}(b).

\begin{figure}[!t]
	\centering
	\includegraphics[width=0.7\columnwidth]{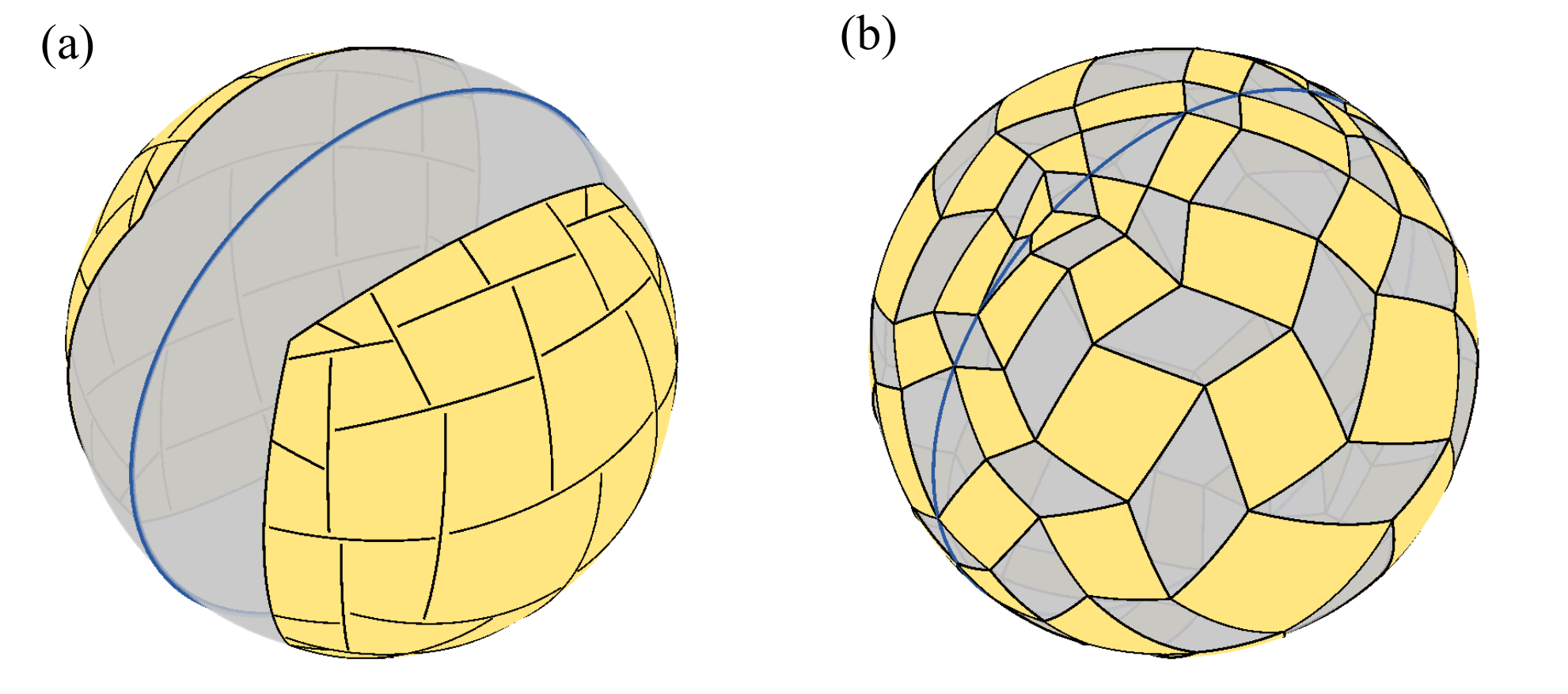}
	\caption{Two mirror-symmetrical reconfigurable SQK shell structures optimized to approximate hemispheres can be connected to form a full sphere. (a) Each kirigami pattern is perforated on a spherical square of side length $s=0.45\pi$ and Gaussian curvature $K=1$. (b) Each deployed configuration covers a hemispherical dome of height $h=1.0$, and is connected to compose a full sphere.}
	\label{fig:SuppHemisphere}
\end{figure}

For the demonstration in Fig.~4 in the Main Text, the input of the optimization is a $6\times6$ square SQK tessellation of side length $s=0.465\pi$ and Gaussian curvature $K=1$. 
The initial tessellation is set to be compatible at $\omega^\star=0.6\pi$, which is obtained from Eqs.~(\ref{eq:supp-sp-EqSys-1})--(\ref{eq:supp-sp-EqSysBC}) and (\ref{eq:supp-sp-opt-loop}) with constant cut ratios $k^b_{i,j}=k^d_{i,j}=0.5$ and uniformly distributed boundary vertices. 
The height of the target spherical dome is $h=1.2$, such that the equation of the boundary circle is $\tilde{g}(\textbf{{r}},\lambda_0)=x-\lambda_0$ with $\lambda_0=-0.2$.
As an additional demonstration, we optimize to obtain a square SQK shell with side length $s=0.45\pi$ that approximates a hemispherical dome ($\lambda_0=0$) at $\omega^\star = 0.7\pi$.
Then we can connect the designed structure and its counterpart with mirror symmetry to achieve a full deployed sphere, as shown in Fig.~\ref{fig:SuppHemisphere}.

\begin{figure}[!t]
	\centering
	\includegraphics[width=1\columnwidth]{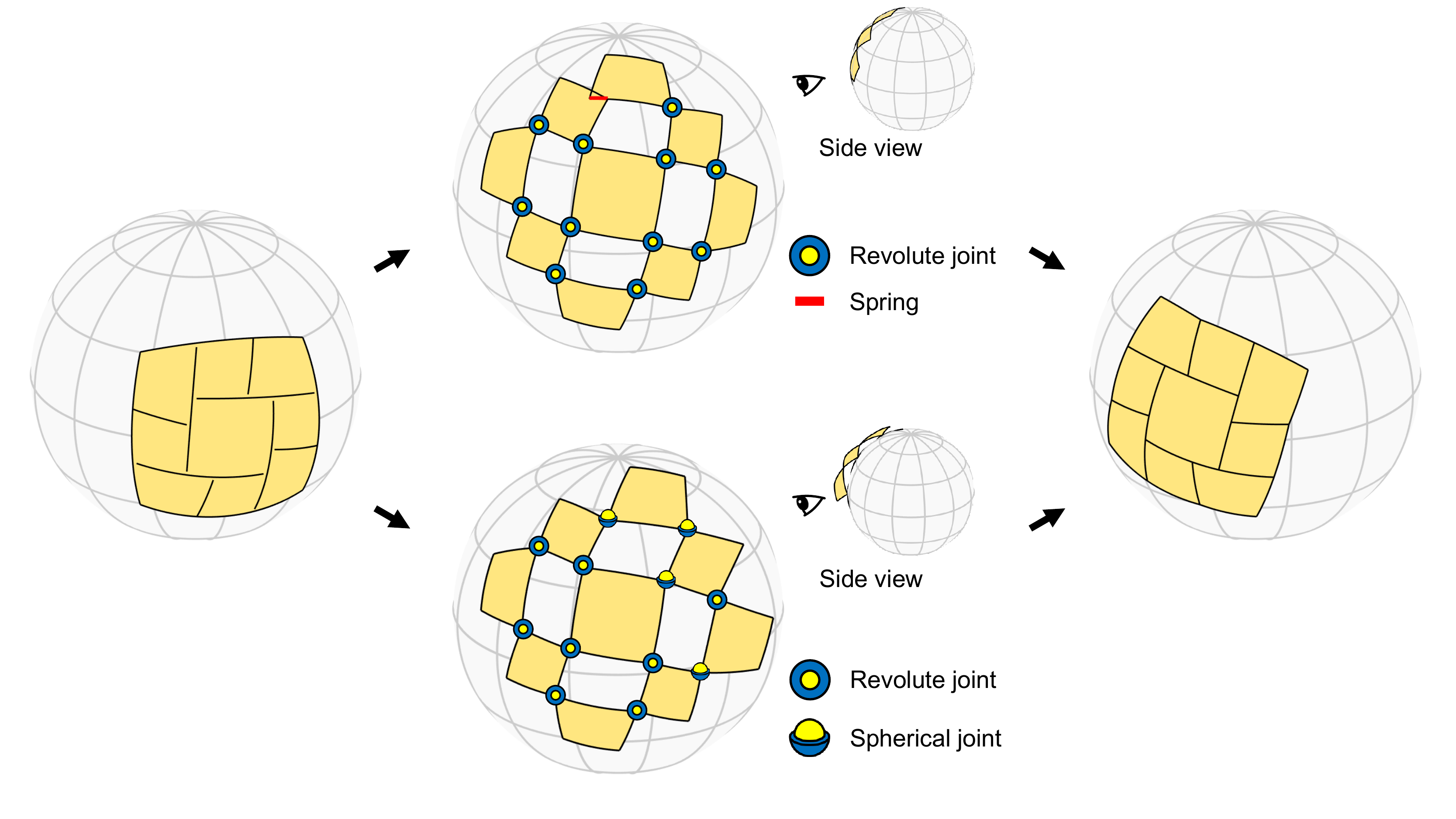}
	\caption{The apparent rigidity of the two-configuration result will vanish if we introduce additional freedom to move in the radial direction. Middle top: the single-spring model incorporates revolute joints and a spring to achieve the transition between two compatible configurations. Middle bottom: spherical joints can be used to permit some panels to move out of the spherical surface, such that zero-energy transition can be realized.}
	\label{fig:SuppSphericalJoint}
\end{figure}

\section{The Role of Out-of-surface Motion}

In the main text, we use the single-spring model and the multispring model to simulate the transition between two compatible configurations of the SQK tessellations.
Essentially, both of these two models introduce elastic energy at the incompatible states to make the transition happen, during which the panels can only move on the spherical surface.
As an realization, the panels can be connected by revolute joints that only permit the rotation around the radial direction of the spherical surface.
Here we supplement a mechanism model that replaces some of the revolute joints by spherical joints to realize zero-energy transition between two compatible configurations. The difference is that the spherical joints allow the panels to have deformations out of the spherical surface.
As shown in Fig.~\ref{fig:SuppSphericalJoint}, the three panels on the right of the kirigami tessellation can move out of the spherical surface by adding four spherical joints, and the tessellation becomes a floppy mechanism with additional degrees of freedom of motion.
In this case, the incompatibility of SQK tessellations are characterized by the out-of-surface motion instead of rigidity.
For SQK tessellations composed of more panels, it is possible to replace carefully-chosen revolute joints by spherical joints to systematically design such mechanisms of SQK tessellations with out-of-surface motion.
We leave this as future work.

\section{Fabrication of Physical Models}

The $5\times5$ SQK tessellation [shown in Figs.~1(c) and 1(d) in the Main Text] is fabricated by the 3D printer Stratasys Objet350 Connex3 with the material TangoBlack. The panels are connected by the rubber-like materials at the joints.

The $6\times6$ SQK dome [shown in Figs.~4(c) and 4(d) in the Main Text] is fabricated by the 3D printer ZRapid iSLA1900D with the material 9400 Resin. The panels are connected by revolute joints.

\section{Movie Caption}

{\bf Movie 1} {\it Configuration transition of the bistable spherical kirigami domelike structure.}